\documentclass[twocolumn]{aastex631}

\usepackage{amsmath}
\usepackage{array}
\usepackage{tabularx}
\usepackage{graphicx}
\usepackage{xcolor}

\begin{document}

\title{Formation of a Protostellar Multiple System via Rotational Fragmentation}

\author[0001-0002-5286-2564]{Tie Liu}\thanks{E-mail: liutie@shao.ac.cn}
\affiliation{State Key Laboratory of Radio Astronomy and Technology, Shanghai Astronomical Observatory, Chinese Academy of Sciences, \\
80 Nandan Road, Shanghai 200030, People's Republic of China}

\author[0000-0003-4506-3171]{Qiuyi Luo}
\affiliation{Department of Astronomy, School of Science, The University of Tokyo, 7-3-1 Hongo, Bunkyo, Tokyo 113-0033, Japan}
\affiliation{Center for Astrophysics, Harvard Smithsonian, 60 Garden Street, Cambridge, MA 02138, USA}

\author[0000-0002-9836-0279]{Siju Zhang}
\affiliation{Departamento de Astronom\'{i}a, Universidad de Chile, Las Condes, 7591245 Santiago, Chile}
\affiliation{Chinese Academy of Sciences South America Center for Astronomy, National Astronomical Observatories, Chinese Academy of
Sciences, Beijing, 100101, China}

\author[0000-0002-2826-1902]{Qilao Gu}
\affiliation{State Key Laboratory of Radio Astronomy and Technology, Shanghai Astronomical Observatory, Chinese Academy of Sciences, \\
80 Nandan Road, Shanghai 200030, People's Republic of China}

\begin{abstract}
We present a multi-scale analysis of the dense core G205.46-14.56-N2 and its host filament G205.46-14.56 using ALMA, Herschel, JCMT, and PMO observations. The filament exhibits a hierarchical fragmentation process primarily governed by thermal Jeans instability. The central region of the dense core G205.46-14.56-N2 hosts a remarkable mirror-symmetric twin binary protostellar system. We detect well-collimated, aligned outflows from all four protostars. Velocity fields traced by H$_2$CO emission reveal clear gradients, and the ratio of rotational kinetic energy to gravitational energy increases with spatial resolution, indicating fast differential rotation within the core. The morphology and kinematics of the quadruple system bear striking resemblance to pure hydrodynamic simulations of rapidly rotating core collapse. These findings (ordered fragmentation and aligned outflows) are inconsistent with the stochastic expectations of turbulent fragmentation and instead may provide direct observational evidence that rotation-driven fragmentation is a viable pathway for forming compact protostellar multiple systems. To our knowledge, this study presents the first high-order (N$\geq$4) protostellar multiple system whose formation can be attributed to rotational fragmentation.
\end{abstract}


\section{Introduction} \label{sec:intro}

 Stars predominantly form in clustered and multiple systems rather than in isolation. Approximately half of all Sun-like stars possess at least one stellar companion, and the multiplicity fraction increases substantially for more massive stars \citep{Duchene2013}. Early numerical simulations indicate that protostellar multiple systems can form via rotational fragmentation of rapidly rotating cores \citep{1991Natur.351..298B,2018MNRAS.478.5460R,2023A&A...673A.134M}. The binary protostellar system L1551 IRS 5 has been proposed to form via rotationally driven fragmentation of its parent core \citep{2016ApJ...826..153L}. However, observational evidence for rotational fragmentation remains scarce. Consequently, current star formation theories favor turbulence over rotation as the primary driver of density fluctuations within cores, which naturally leads to fragmentation and multiple star formation \citep[see review by][]{2023ASPC..534..275O}. High-resolution interferometric surveys using the Jansky Very Large Array (JVLA) and the Atacama Large Millimeter/submillimeter Array (ALMA) have provided observational support for the turbulent fragmentation scenario in the formation of binary and multiple protostellar systems \citep{Pineda2015,Tobin2016a,Tobin2016b,LeeJE2017,2022ApJ...931..158L}. Nevertheless, the physical mechanisms governing the formation and early evolution of multiple stellar systems remain poorly constrained.

Recent observations of the dense molecular core G205.46-14.56-N2 (also known as SSV 63) in the Orion B molecular cloud at a distance of $404\pm4$ pc have revealed an exceptional protostellar multiple system hosting at least seven components, with masses ranging from the hydrogen-burning limit up to proto-Herbig Ae stars \citep{2022ApJ...931..158L,2023AJ....165..209R}. This system harbors five collimated jets, and five of its embedded protostellar members are surrounded by small disks (sizes: 70 to 320 au), as resolved in 1.3 mm continuum emission with ALMA \citep{2023AJ....165..209R}. In previous studies, a wide-separation quadruple system (3300--11400 au), composed of a young protostar and three gravitationally bound dense gas condensations, was reported in a filament of the Barnard 5 region \citep{Pineda2015}. However, it remains unclear whether the three dense gas condensations in that system will eventually form protostars. In contrast, all five embedded members of G205.46-14.56-N2 have already formed protostars \citep{2023AJ....165..209R}, making this multiple system a uniquely valuable laboratory for investigating the formation and early dynamical evolution of compact protostellar clusters.

\cite{2023AJ....165..209R} have presented a detailed analysis of the core, jets/outflows, and disks of this multiple protostellar system but did not discuss its formation mechanism. In this work, we perform a multi-wavelength, multi-scale analysis of the dense core G205.46-14.56-N2 and its host filament (G205.46-14.56) to determine the dominant formation mechanism of this compact protostellar multiple system.

\section{Observations} \label{sec:obs}

\subsection{ALMA observations} \label{Obs:ALMA}

The ALMA observations of G205.46-14.56-N2 were carried out as part of the ALMASOP project (ID: 2018.1.00302.S.; PI: Tie Liu) with ALMA Band 6 in Cycle 6, during October 2018 to January 2019. The observations were executed in four blocks using three different array configurations: 12 m C43-5 (TM1), 12 m C43-2 (TM2), and the 7 m ACA. For observations in the C43-5, C43-2, and compact 7 m ACA configurations, the unprojected baseline lengths range from 15 to 1398 m, 15 to 500 m, and 9 to 49 m, respectively. The resulting maximum recoverable scale is 25$^{\prime\prime}$. The ALMA Band 6 receivers were used to simultaneously capture four spectral windows (SPWs), each with a total bandwidth of 1.875 GHz and a velocity resolution of approximately 1.3 km~s$^{-1}$. The synthesized beam size and 1$\sigma$ rms of the continuum emission from combining all configuration data are $0^{\prime\prime}.33\times0^{\prime\prime}.29$ (P.A. = $-66.3^\circ$) and $\sim0.1$ mJy~beam$^{-1}$, respectively. The synthesized beam size and 1$\sigma$ rms of the continuum emission from the ACA configuration data alone are $7^{\prime\prime}.5\times4^{\prime\prime}.1$ (P.A. = $-82.1^\circ$) and $\sim2$ mJy~beam$^{-1}$, respectively. The synthesized beam size and 1$\sigma$ rms of the continuum emission from the combined ACA and TM2 configuration data are $1^{\prime\prime}.4\times1^{\prime\prime}.1$ (P.A. = $-63.7^\circ$) and $\sim0.35$ mJy~beam$^{-1}$, respectively. Further details of the ALMA observations can be found in the ALMASOP survey description paper {\citep{Dutta2020}.

\subsection{Archival infrared data} \label{Obs:infra}
We have made use of data from the Herschel Gould Belt Survey (HGBS) project \citep{Andre2010}. We also used the dust temperature and column density maps derived from \textit{Herschel} data {\citep{Konyves2020}, which were smoothed to the resolution of the 500 $\mu$m map (36$^{\prime\prime}.3$). Additionally, we used \textit{Spitzer} IRAC archival data.

\subsection{JCMT data} \label{Obs:JCMT}
The James Clerk Maxwell Telescope (JCMT) observations of G205.46-14.56 were conducted as part of the JCMT Transient Survey \citep{Mairs2017,Herczeg2017}. We used the co-added 850 $\mu$m map, which has an angular resolution of 14$^{\prime\prime}.6$ and an rms level of $\sim4$ mJy~beam$^{-1}$.

\subsection{PMO observations} \label{Obs:PMO}
Mapping observations of G205.46-14.56 in the $^{13}$CO (1-0) line were conducted using the PMO 13.7 m radio telescope on September 9, 2020. The nine-beam array receiver system in single-sideband (SSB) mode was used as the front end. FFTS spectrometers were used as back ends, providing a total bandwidth of 1 GHz and 16384 channels, corresponding to a velocity resolution of 0.17 km~s$^{-1}$ for $^{13}$CO (1-0). The half-power beam width is 56$^{\prime\prime}$, and the main beam efficiency is $\sim0.5$. The pointing accuracy of the telescope was better than 4$^{\prime\prime}$. The typical system temperature ($T_{\rm sys}$) in SSB mode is approximately 110 K, varying by about 10\% for each beam. The on-the-fly (OTF) observing mode was employed. The antenna continuously scanned a region of $30^{\prime}\times30^{\prime}$ at a scan speed of 20$^{\prime\prime}$~s$^{-1}$. The typical rms noise level was 0.1 K. Using the GILDAS software package (including CLASS and GREG), the OTF data were converted to three-dimensional cubes with a grid spacing of 30$^{\prime\prime}$, and the baselines were corrected by fitting linear functions.

\section{Results} \label{sec:results}

\subsection{Hierarchical fragmentation of the filament G205.46-14.56}

\begin{figure*}[ht]
\centering
\includegraphics[width=\linewidth]{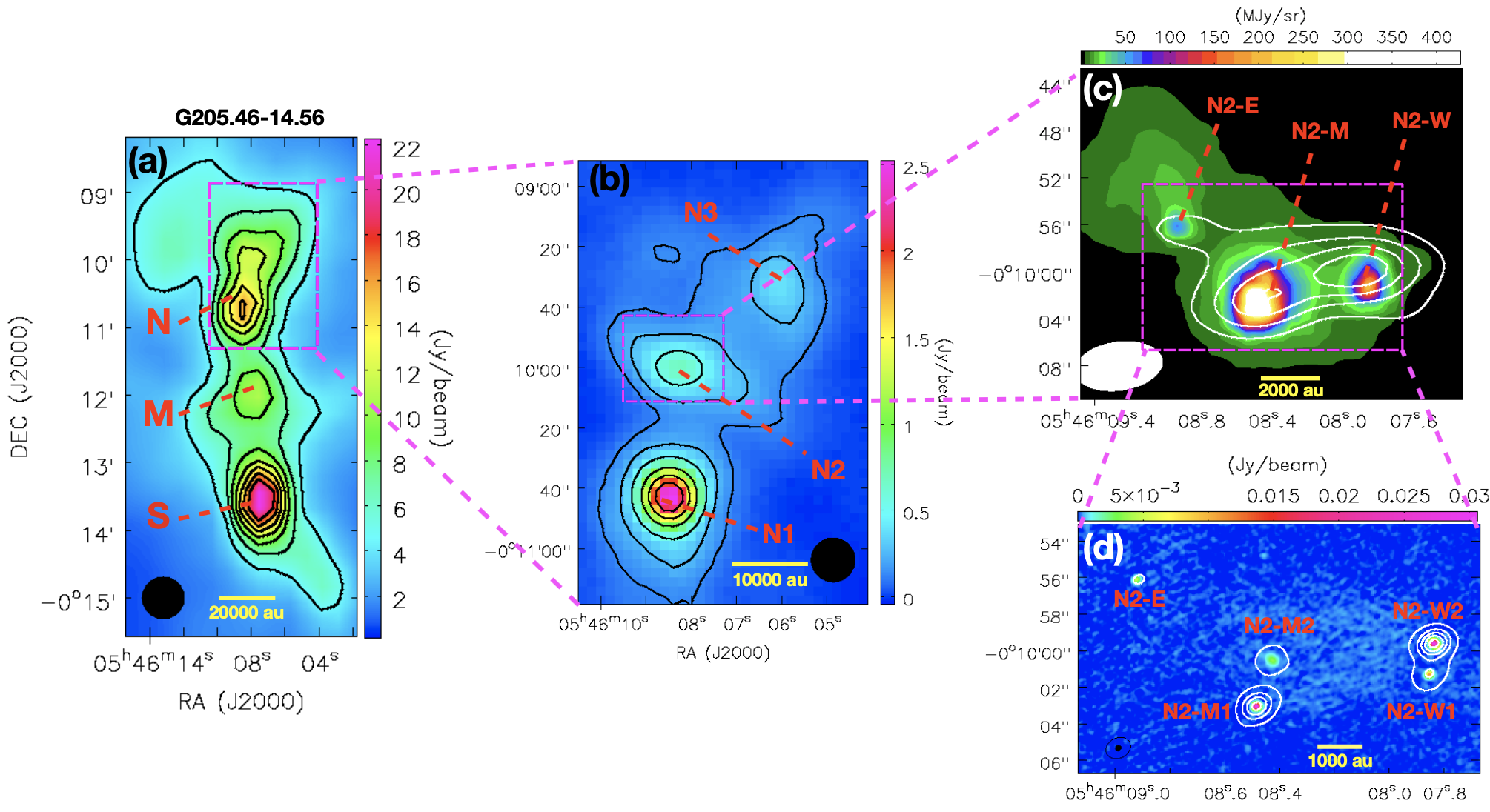}
\caption{Hierarchical fragmentation of the G205.46-14.56 filament. (a) Herschel 500 $\mu$m continuum emission. Contours are shown from 10\% to 90\% in steps of 10\% of the peak flux intensity (22.3 Jy~beam$^{-1}$). (b) JCMT 850 $\mu$m continuum emission. Contours are [0.05, 0.1, 0.2, 0.4, 0.6, 0.8] times the peak flux intensity (2.53 Jy~beam$^{-1}$). (c) 1.3 mm continuum emission from ACA 7-m array observations is shown in contours overlaid on Spitzer 4.5 $\mu$m emission. Contours are [0.2, 0.4, 0.6, 0.8]$\times$86.4 mJy~beam$^{-1}$. (d) Contours show 1.3 mm continuum emission from combined ACA and ALMA data with an angular resolution of $\sim1^{\prime\prime}$. Contour levels are [0.1, 0.3, 0.5, 0.7, 0.9]$\times$42.7 mJy~beam$^{-1}$. The background image shows 1.3 mm continuum emission from the combined ACA and ALMA data with an angular resolution of $\sim0^{\prime\prime}.35$.}
\label{fig1:fragmentation}
\end{figure*}

Fig.~\ref{fig1:fragmentation} presents the multi-scale images of the filament G205.46-14.56 obtained from Herschel, JCMT and ALMA observations, achieving spatial resolutions from $\sim$140 au to 0.07 pc. Dense structures in these images were extracted using the \textit{Astrodendro} algorithm \citep{Rosolowsky2008}. The details can be found in Appendix \ref{sec:frag}. Their physical parameters including radii, masses and separations are calculated in Appendix \ref{sec:frag}, and presented in Table~\ref{tab:frag} and Table~\ref{tab:MST}.

The filament G205.46-14.56 has a length of $\sim0.7$ pc and a mass of $\sim30 M_{\odot}$. We find that this filament undergoes hierarchical fragmentation, a phenomenon also seen in infrared dark clouds \citep[e.g.,][]{2011ApJ...735...64W,2014MNRAS.439.3275W}. As illustrated in the Herschel/SPIRE 500 $\mu$m image (beam $\sim36''$; corresponding to a spatial resolution of $\sim14440$ au; Fig.~\ref{fig1:fragmentation}a), the filament consists of three gas clumps (N, M, S) with a nearly uniform separation of $\sim39000$ au. Higher resolution 850 $\mu$m emission data from the  (JCMT; beam $\sim14.6''$; $\sim5900$ au spatial resolution) reveal that the northern clump (G205.46-14.56-N) further fragments into three dense cores (N1, N2, N3; Fig.~\ref{fig1:fragmentation}b), with a nearly uniform separation of $\sim17000$ au. The central core (G205.46-14.56-N2) is highly flattened, with a radius of $\sim7000$ au and a gas mass of $\sim2.3\pm0.7~M_{\odot}$. Virial analysis indicates that this flattened core is likely gravitationally bound (Appendix \ref{sec:virial}). Observations from the ALMA Compact Array (ACA) 7-m array (beam $\sim4''$; $\sim1600$ au spatial resolution) show that N2 fragments into three condensations (N2-E, N2-M, N2-W; Fig.~\ref{fig1:fragmentation}c), with a mean separation of $\sim3200$ au. At the highest spatial resolution achieved with ALMA observations (beam $\sim0.35''$; $\sim140$ au; Fig.~\ref{fig1:fragmentation}d), the N2-E condensation does not undergo further fragmentation. In contrast, the N2-M and N2-W condensations fragment further, each forming two protostars with separations of $\sim1080$ au and $\sim700$ au, respectively. 

Following the MST-based analysis of filament fragmentation presented by \citet{Clarke2019}, we constructed a minimum spanning tree \citep[MST;][]{Gower1969} for the identified structures and analyzed the projected separations between structures connected by the MST. Table~\ref{tab:MST} lists, for each fragmentation level, the calculated projected separation and the corresponding thermal Jeans length of the parental structures. The results demonstrate that each level fragments into nearly equally spaced cores, with the observed separation closely matching the thermal Jeans length of the parental structure considering projection. This agreement suggests that the hierarchical fragmentation observed in filament G205.46-14.5 is dominated by thermal Jeans instability, in contrast to the turbulence-controlled fragmentation found in infrared dark clouds \citep[e.g.,][]{2011ApJ...735...64W,2014MNRAS.439.3275W}.

\subsection{The central twin binary system in the G205.46-14.56-N2 core}

\begin{figure*}[ht]
\centering
\includegraphics[scale=0.2]{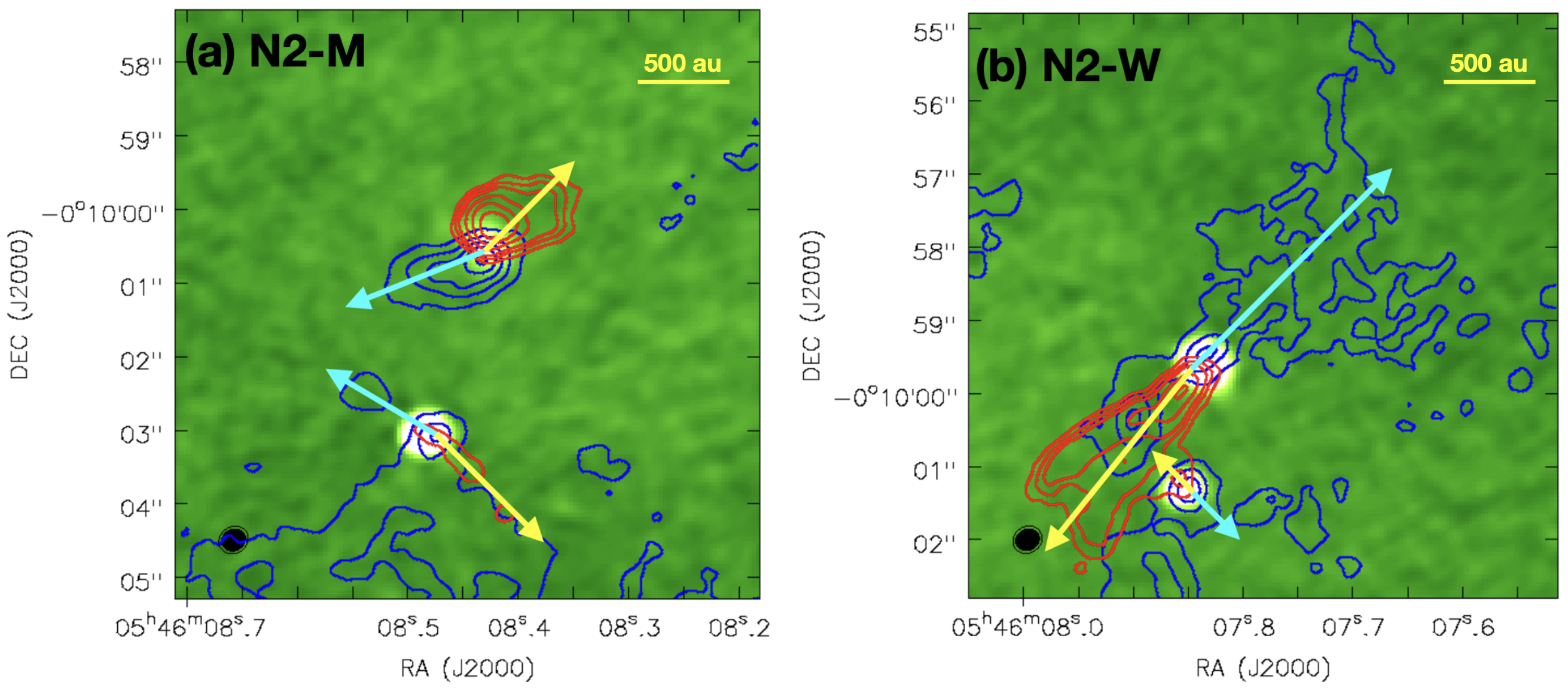}
\caption{Molecular CO outflows are shown in contours, and 1.3 mm continuum is shown in color image. Red and blue contours represent redshifted (from 15 to 35 km~s$^{-1}$) emission and blueshifted from -15 to 5 km~s$^{-1}$) emission, respectively.  Contour levels are [1,2,3,4,5]$\times0.1$ Jy~beam$^{-1}$~km~s$^{-1}$ for blueshifted emission and [1,1.5,2,3,4,5]$\times0.1$ Jy~beam$^{-1}$~km~s$^{-1}$ for redshifted emission. Arrows mark outflow directions determined by visual identification. }
\label{fig:outflows}
\end{figure*}

\begin{figure}[ht]
\centering
\includegraphics[scale=0.15]{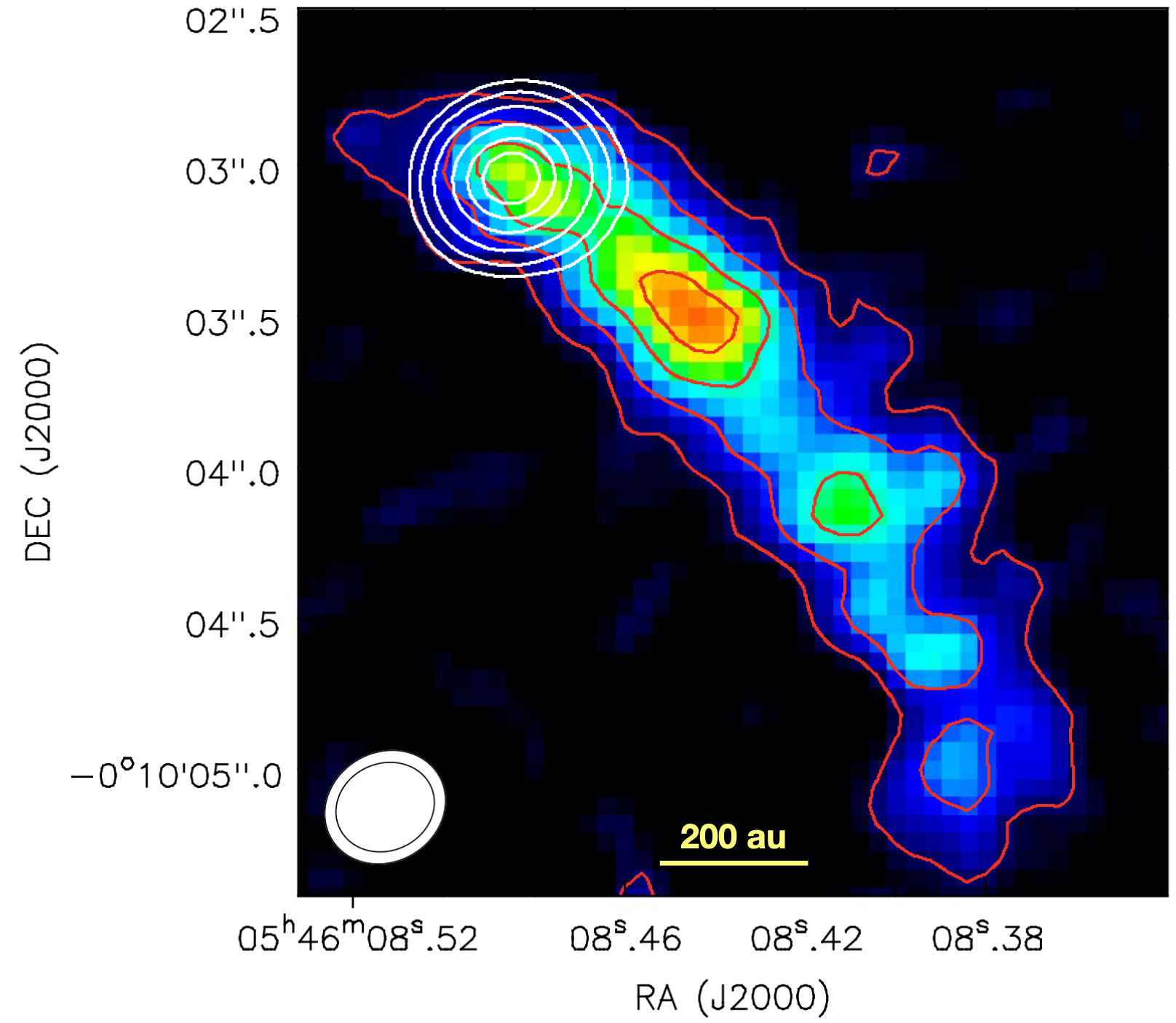}
\caption{Close-up view of the redshifted CO outflow of N2-M1, rendered in color scale with red contours. The red contour levels are [0.3,0.5,0.7,0.9]$\times0.152$ Jy~beam$^{-1}$~km~s$^{-1}$. The 1.3 mm continuum is shown in white contours. Contour levels are [0.05, 0.1, 0.2, 0.4, 0.6, 0.8]$\times0.307$ Jy~beam$^{-1}$.}
\label{fig:N2M1red}
\end{figure}

\begin{figure*}[ht]
\centering
\includegraphics[scale=0.23]{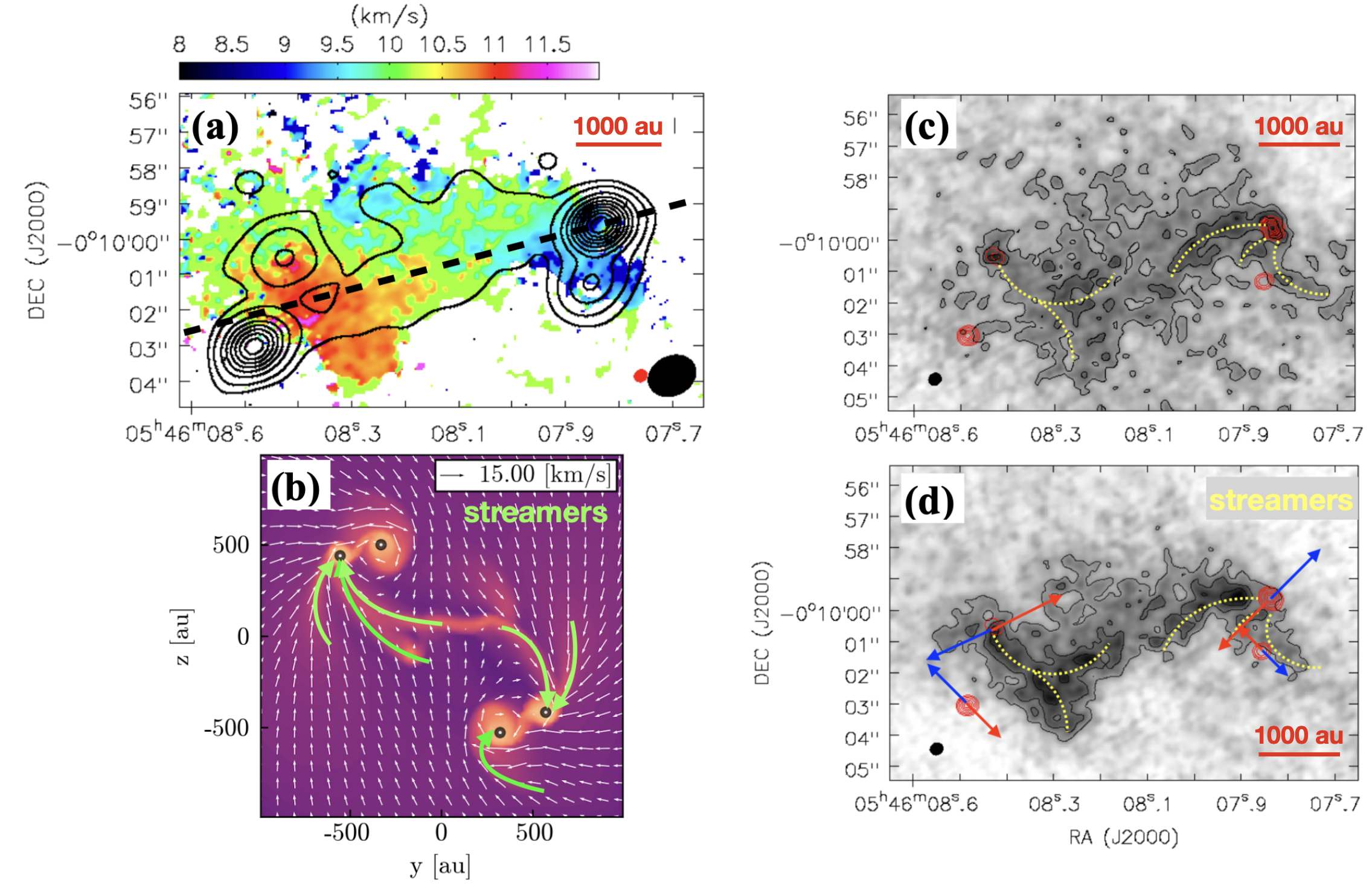}
\caption{(a) The mirror-symmetric twin binary protostellar systems. Color image shows the velocity field of H$_2$CO line emission. Contours show the 1.3 mm continuum emission obtained from ACA+TM2 data. Contour levels are [0.05, 0.06, 0.07, 0.08, 0.09, 0.1, 0.2, 0.4, 0.6, 0.8]$\times$42.7 mJy~beam$^{-1}$. The elongated bar-like structure is outlined by the dashed line. (b) Column density of a fragmented dense core in the purely hydrodynamical model \citep{2023A&A...673A.134M}. White dots indicate sink particle positions. Gas velocity vectors are overplotted. This figure is adapted from \cite{2023A&A...673A.134M} with permission from the author. (c) The gray image and black contours show the integrated intensity map of C$^{18}$O (2-1) from 6 to 14 km~s$^{-1}$. Contour levels are [0.4, 0.6, 0.8]$\times$0.203 Jy~beam$^{-1}$~km~s$^{-1}$. (d) The gray image and black contours show the integrated intensity map of H$_2$CO 3(0,3)-2(0,2) from 8 to 12 km~s$^{-1}$. Contour levels are [0.4, 0.6, 0.8]$\times$0.1245 Jy~beam$^{-1}$~km~s$^{-1}$. Blue and red arrows indicate the directions of CO outflows. Yellow dashed lines in panels (c) and (d) mark the streamer-like gas structures, and red contours show the 1.3 mm continuum from combined ACA and ALMA data with an angular resolution of $\sim0^{\prime\prime}.35$. Contour levels are [0.05, 0.1, 0.2, 0.4, 0.6, 0.8]$\times$0.0307 Jy~beam$^{-1}$.}
\label{fig:rotate}
\end{figure*}

Remarkably, the four deeply embedded protostellar members (N2-M1, N2-M2, N2-W1, N2-W2) in the central region constitute a \textit{mirror-symmetric twin binary protostellar system} (Fig.~\ref{fig1:fragmentation}d). These twin binaries exhibit nearly identical separations, evolutionary stages, and even aligned outflow orientations (Fig.~\ref{fig:outflows}). Such ordered, symmetric fragmentation and their aligned collimated outflows are inconsistent with predictions from the turbulent fragmentation paradigm, in which angular momentum distribution and fragmentation outcomes are expected to be stochastic \citep{Offner2010,LeeJE2017}.

Utilizing higher-resolution ALMA CO J=2-1 molecular line data compared to the dataset published in \cite{2023AJ....165..209R}, we visually identify protostellar outflows from the high-velocity integrated intensity map presented in Fig.~\ref{fig:outflows}. These outflows display mirror-symmetric orientations (Fig.~\ref{fig:outflows}). We detect faint outflows toward N2-M1 and N2-W1. Notably, the blob-shaped blueshifted lobes of N2-M1 and N2-W1 are faint and difficult to disentangle from surrounding cloud material. Accordingly, their derived properties must be interpreted with caution. Additionally, the redshifted lobe of N2-W1 is highly compact and lies in close proximity to the redshifted lobe of N2-W2. In contrast, the redshifted lobe of N2-M1 shows prominent jet-like structure threaded by a linear chain of knots (Fig.~\ref{fig:N2M1red}), a feature that was not identified in the earlier analysis of \cite{2023AJ....165..209R}. The outflows driven by N2-M2 and N2-W2 are much stronger than those of N2-M1 and N2-W1. The outflows driven by N2-M2 and N2-W2 have similar orientations along the northwest-southeast direction, while the outflows of N2-M1 and N2-W1 are oriented along a perpendicular direction (northeast-southwest). These aligned outflows could be attributed to orbital motions within the twin binary system. The outflow parameters are derived in Appendix \ref{sec:outflows}. The outflows are dynamically young, with estimated dynamical ages ranging from $\sim130$ yr to $\sim860$ yr.  The accretion rates inferred from the outflows for N2-M2 and N2-W2 are approximately one order of magnitude larger than those for N2-M1 and N2-W1. This indicates that N2-M2 and N2-W2 are at younger evolutionary stages and still undergoing active gas accretion. 

We identify clear velocity gradients within the central quadruple system (the twin binaries), as traced by the velocity field of the $\text{H}_2\text{CO}$ line emission (Fig.~\ref{fig:rotate}a and Fig.~\ref{fig:moment1}), hinting at kinematic signatures of rapid rotation (Appendix \ref{sec:rotation}). The velocity gradients, or the ratio of rotational kinetic energy to gravitational energy ($\beta$), increase with spatial resolution (Appendix \ref{sec:rotation}), indicating differential rotation of the system. Strikingly, the morphology and kinematics of the quadruple system bear a remarkable resemblance to pure hydrodynamical simulations of rapidly rotating core collapse \citep[Fig.~\ref{fig:rotate}b; ][]{1991Natur.351..298B,2023A&A...673A.134M}. In these simulations, ordered multiple systems form via the hydrodynamical collapse of a rapidly rotating dense core \citep[e.g., ][]{1991Natur.351..298B,2023A&A...673A.134M}. Thus, the ALMA observations may provide direct evidence that rotation-driven fragmentation is a viable pathway for forming compact multiple protostellar systems.

Using high-resolution ALMA molecular line data, we even resolve finer gas structures within the core. The twin binary components are connected by a prominent bar-like or bridge-like elongated gaseous structure (hereafter referred to as ``the bar''), clearly detected in both 1.3 mm continuum emission and molecular C$^{18}$O and H$_2$CO line emission (Fig.~\ref{fig:rotate}). Furthermore, this bar exhibits streamer-like substructures, highlighted by the yellow dashed lines in Fig.~\ref{fig:rotate}c and Fig.~\ref{fig:rotate}d. These streamer-like substructures likely trace gas accretion flows toward the disks, as also observed in numerical simulations \citep[see Fig.~\ref{fig:rotate}b; ][]{2023A&A...673A.134M}.

\section{Discussions and Summary} \label{sec:sum}

Hierarchical multiple systems similar to that in G205.46-14.56-N2 can form via the hydrodynamical collapse of rapidly rotating cores in pure hydrodynamical simulations \citep{1991Natur.351..298B,2018MNRAS.478.5460R,2023A&A...673A.134M}. To the best of our knowledge, this study presents the first robustly reported high-order (N$\geq$4) protostellar multiple system whose formation can be attributed to rotational fragmentation. Magnetohydrodynamical (MHD) simulations suggest that magnetic fields can strongly affect core fragmentation or limit the growth of protostellar disks \citep[e.g., ][]{2014ApJ...794...44B,2023A&A...673A.134M}. Notably, the observed central quadruple protostellar system (Fig.~\ref{fig:rotate}a) shows striking morphological similarity to the multiple systems formed in pure hydrodynamical simulations \citep{1991Natur.351..298B,2023A&A...673A.134M} but differs substantially from the outcomes of non-ideal radiation-MHD models \citep{2023A&A...673A.134M}. This comparison tentatively suggests that magnetic fields may play a subdominant role during the fragmentation and disk formation process of the hierarchically structured protostellar multiple system in G205.46-14.56-N2.

Finally, the dense filament G205.46-14.56, which hosts this core, is also rotating with $\beta\sim0.04$ \citep{Hsieh2021}. Moreover, the molecular cloud that contains this filament exhibits a significant large-scale velocity gradient (Appendix \ref{cloud}), suggesting that large-scale convergent flow or cloud-cloud collision may play a role in the formation of the filament itself. 

In summary, this study emphasizes the importance of rotation in the hierarchical fragmentation and formation of protostellar multiple systems. Specifically, cloud-cloud collisions or large-scale convergent flows within molecular clouds give rise to dense, rotating filaments, which then undergo hierarchical fragmentation to produce rapidly rotating dense cores. Rotation within these cores subsequently triggers gravitational instability and fragmentation, ultimately leading to the formation of binary or multiple star systems. This paradigm of multiple star system formation awaits further scrutiny through more multi-scale observational studies and higher-fidelity numerical simulations in the future.

\clearpage

\section*{Acknowledgments}
T.L. acknowledges support from the National Science and Technology Major Project of China (No. 2024ZD1100601), the National Key R\&D Program of China (No. 2022YFA1603100), the National Natural Science Foundation of China (NSFC) through grants No. 12073061 and No. 12122307, the Tianchi Talent Program of Xinjiang Uygur Autonomous Region, and the Tianshan Talent Training Program (2024TSYCTD0013). 

This paper makes use of the following ALMA data: ADS/JAO.ALMA\#2018.1.00302.S. ALMA is a partnership of ESO (representing its member states), NSF (USA), and NINS (Japan), together with NRC (Canada), NSC, and ASIAA (Taiwan), as well as KASI (Republic of Korea), in cooperation with the Republic of Chile. The Joint ALMA Observatory is operated by ESO, AUI/NRAO, and NAOJ.

\clearpage

\appendix

\renewcommand\thefigure{\Alph{section}\arabic{figure}}
\renewcommand\thetable{\Alph{section}\arabic{table}}

\section{Fragmentation analysis} \label{sec:frag}
Dense structures were extracted using the \textit{Astrodendro} algorithm \citep{Rosolowsky2008}. These dense structures correspond to the ``leaves'' in the dendrogram of each image, meaning that no substructure is embedded within them. Three key parameters need to be set in \textit{Astrodendro}: (1) \texttt{min\_value}, the minimum value included in the dendrogram, set to 5$\sigma$; (2) \texttt{min\_delta}, the threshold for differentiating leaves, set to 1$\sigma$; and (3) \texttt{min\_npix}, the minimum size considered as an independent leaf, set to one synthesized beam. To validate the robustness of our derived structures, we systematically adjust key parameters within \textit{Astrodendro}. As a representative test, we alter the \textit{min\_value} threshold from 4$\sigma$ up to 6$\sigma$. This variation barely alters our outputs, and the identified structures are fully consistent with our visual checks. Our \textit{Astrodendro} structure extraction on this filament yields robust measurements, as the fragments resolved at every hierarchical level are well separated by no less than one beam width (see Fig.~\ref{fig1:fragmentation}). The derived properties of the dense structures are shown in Table~\ref{tab:frag}.

\textit{Astrodendro} may fail to identify leaves in crowded environments \citep{Kong2021}. To mitigate this issue, we required the dendro-identified structures to have a major-to-minor axis ratio $<2$. For elongated structures with a major-to-minor axis ratio $>2$, we searched for sub-peaks within the structure's area and then performed 2D Gaussian fitting using the \textit{CASA} task \texttt{imfit}. Only one elongated structure was extracted, corresponding to the main emission structure in the ACA 1.3 mm image. Using the \textit{Spitzer} 4.5 $\mu$m emission as a reference, three 2D Gaussians were fitted to this elongated structure.

\textit{Astrodendro} derives a radius $R_{\rm dendro}$ (rms radius) using the intensity-weighted second moment. After deconvolution, $R_{\rm dendro}$ is generally smaller than the separations between embedded smaller condensations in our cases; therefore, it is not suitable for fragmentation analysis. We instead use the effective radius $R_{\rm eff} = \sqrt{Area/\pi}$, where $Area$ is the size of the leaf region extracted by \textit{Astrodendro} \citep{Krieger2020, Hatchfield2020, Lewis2021, Kong2021, Phiri2021, Takemura2021}. The beam-deconvolved effective radius is $R_{\rm eff-de} = \sqrt{{R_{\rm eff}}^2 - ({\rm Beam_{\rm FWHM}}/2)^2}$ \citep{Lewis2021}. Previous studies have shown that $R_{\rm eff-de}$ is approximately $2R_{\rm dendro}$ for well-resolved structures \citep{Krieger2020, Phiri2021}. 

Assuming the dust continuum emission is optically thin, the mass of each structure is estimated following \citep{Hildebrand1983}:
\begin{equation}
\centering
M = R_{\rm gd} \frac{F_{\nu}D^{2}}{\kappa_{\nu}{B_{\nu}(T_{\rm dust})}}, 
\label{MassEquation}
\end{equation}
where $F_{\nu}$, $\kappa_{\nu}$, and $B_{\nu}(T_{\rm dust})$ are the measured flux, dust opacity per gram, and the Planck function at frequency $\nu$, respectively. The gas-to-dust mass ratio $R_{\rm gd}$ is assumed to be 100 in this work. $\kappa_{\nu}$ is set to 0.9, 1.5, and 3.9 cm$^{2}$~g$^{-1}$ for ALMA (ACA) 1.3 mm, JCMT 850 $\mu$m, and Herschel 500 $\mu$m, respectively, corresponding to the opacity of dust grains with thin ice mantles at gas densities of $\sim10^{6}$ cm$^{-3}$ (for ALMA and ACA) and $\sim10^{5}$ cm$^{-3}$ (for JCMT and Herschel) \citep{Ossenkopf1994}. 

The thermal Jeans length of parental structure is estimated as:
\begin{equation}    
\lambda_{\rm J}^{\rm th} = \sigma_{\rm th} \left( \frac{\pi}{G\rho} \right) ^{1/2},
\label{equ-lambdaj}
\end{equation}
where $\rho$ is the mass density and $\sigma_{\rm th}$ is the thermal velocity dispersion:
\begin{equation}  
\rho =\frac{M}{\frac{4}{3} \pi {R_{\rm eff-de}}^3}, \quad
\sigma_{\rm th} = \left(\frac{k_{\rm B} T}{\mu m_{\rm H}} \right) ^{1/2}.
\label{equ-thermal}
\end{equation}

The nearest-neighbor separations of adjacent structures are derived using the minimum spanning tree (MST) method \citep{Clarke2019}. The separation and thermal Jeans length calculated at each fragmentation
level are presented in Table \ref{tab:MST}.

\begin{table*}[!thb]
\centering
\caption{Properties of dense structures identified in the G205.46-14.56 filament.}
\renewcommand{\arraystretch}{1.5} 
\begin{tabular}{ccccccccccc}
\hline
\hline
Telescope            & Objects & RA  & DEC                     & Flux$_{\rm peak}$    & Flux$_{\rm total}$  & $R_{\rm eff-de}$ & Mass     & n\tablenotemark{a}     & $T_{\rm dust}$\tablenotemark{b} \\
                     &         &     &                         & (Jy~beam$^{-1}$)     & (Jy)                & (au)         & (M$_{\odot}$)  & (10$^{6}$~cm$^{-3}$)  &    (K)     \\
\hline
Herschel             & N       & 05:46:08.11 &-00:10:18.02   &  16.1$\pm$0.5  & 38.29$\pm$1.77          & 14870 & 13.3$\pm$4.2         & 0.1    &   15          \\
     & M       & 05:46:07.97 &-00:11:57.99  &  10.7$\pm$0.5  & 12.20$\pm$0.66         & 6930   & 5.4$\pm$1.8              & 0.6 &   14          \\
                     & S       & 05:46:07.34 &-00:13:30.08  &  22.3$\pm$0.5  & 35.00$\pm$1.31            & 12240  & 10.6$\pm$3.4      &  0.2    &   16          \\
\hline
JCMT                 & N1      & 05:46:08.48 &-00:10:43.76  & 2.110$\pm$0.004 & 4.040$\pm$0.033         & 8740 & 8.1$\pm$2.6            &  0.4  &   14          \\
           & N2      & 05:46:08.00 &-00:09:59.91  & 0.557$\pm$0.004 & 1.548$\pm$0.023          & 6980 & 2.3$\pm$0.7           & 0.2   &   16           \\
                     & N3      & 05:46:06.08 &-00:09:35.28  & 0.318$\pm$0.004 & 0.688$\pm$0.013          & 4900 & 1.6$\pm$0.5       & 0.5       &   13           \\
\hline
                     &         &     &                         & (mJy~beam$^{-1}$)     & (mJy)                & (au)         & (M$_{\odot}$)  & (10$^{6}$~cm$^{-3}$) &  (K)      \\
\hline
ACA                  & N2-E    & 05:46:08.75   & -00:09:56.50    & 20.7$\pm$3& 45$\pm$25 &  2010  & 0.21$\pm$0.14 &  0.9 & 16 \\
               & N2-M    & 05:46:08.45  & -00:10:02.24  & 70.9$\pm$3 & 77$\pm$17     &  1040  & 0.36$\pm$0.14 & 11.5 & 16 \\
                     & N2-W    & 05:46:07.90  & -00:09:59.83  & 86.4$\pm$3  & 123$\pm$20    &  1340  & 0.59$\pm$0.21 & 8.8 & 16 \\
\hline
ALMA             & N2-E    & 05:46:08.92 & -00:09:56.11  &  4.98$\pm$0.15 &  4.47$\pm$0.28  &  80  & 0.021$\pm$0.007 & 1467.5 
& 16\\
               & N2-M1   & 05:46:08.48 & -00:10:03.04  & 30.29$\pm$0.15 & 33.01$\pm$0.60  &  136 & 0.154$\pm$0.05  & 2190.5 & 16\\
                     & N2-M2   & 05:46:08.43 & -00:10:00.50  &  3.70$\pm$0.15 &  6.95$\pm$0.55  &  129 & 0.032$\pm$0.01 &533.4  & 16\\
                     & N2-W1   & 05:46:07.85 & -00:10:01.35 & 11.40$\pm$0.15 & 11.75$\pm$0.44   &  113 & 0.056$\pm$0.018 & 1388.6  & 16\\
                     & N2-W2   & 05:46:07.84 & -00:09:59.60   & 30.68$\pm$0.15 & 42.54$\pm$0.80 & 162  & 0.2$\pm$0.07  &1683.2   & 16\\
\hline
\end{tabular}
\tablenotemark{a}{Particle number density n and mass density $\rho$ are related as $\rho=\mu m_{H}n$, where $\mu$=2.37.}
\tablenotetext{b}{Dust temperature adopted from Herschel observations \citep{Konyves2020}.}
\label{tab:frag}
\end{table*}

\begin{table*}[!thb]
\centering
\caption{MST separations and Jeans lengths}
\renewcommand{\arraystretch}{1.5} 
\begin{tabular}{cccccccccc}
\hline
\hline
Telescope          & $\lambda_J^{\rm th}$ in parental structure & MST-node   & Separation & Parental Structure \\
                   & (au)                 &            &   (au)     &           \\
\hline
Herschel           &       & N to M          &  40390       & -     \\
Spire 500 $\mu$m   &                      & M to S          &  37400       & -      \\
\hline
JCMT               & 14000      & N1 to N2        &   17950      &  N      \\
850 $\mu$m         & 14000        & N2 to N3        &   15320      &  N      \\
\hline
ACA                & 5700\tablenotemark{a}      & N2-W to N2-M     &   3480      &  N2      \\
1.3 mm             & 5700\tablenotemark{a}      & N2-M to N2-E     &   2940      &  N2     \\
\hline
ACA+ALMA           &  5700\tablenotemark{a}              & N2-E to N2-M2       & 3480  &  N2        \\
1.3 mm             &                & N2-M2 to N2-M1      & 1080  &  N2-M    \\
                   &  5700\tablenotemark{a}              & N2-M2 to N2-W1      & 3490  &  N2      \\
                   &                 & N2-W1 to N2-W2      & 700   & N2-W      \\
\hline
\end{tabular}
\tablenotetext{a}{In calculating the thermal Jeans length of core N2, the mass of N2 is taken to be 8.6 M$_\odot$, which includes both gas and stellar mass.}
\label{tab:MST}
\end{table*}

\section{Virial analysis of the dense molecular core} \label{sec:virial}
The total gas mass of the G205.46-14.56-N2 dense core is $2.3\pm0.7$ M$_{\odot}$. The virial mass of the dense core, assuming a power-law density distribution ($n(r)\propto r^{-p}$), can be derived as:
\begin{equation}  
M_{\rm vir}\approx209\frac{(R/1\ {\rm pc})(\Delta V/1\ {\rm km~s^{-1}})^2}{a_1a_2} M_{\odot},
\label{Mvir}
\end{equation}
where $a_1 = \frac{1-p/3}{1-2p/5}$ is the correction factor for a power-law density distribution, and $a_2$ is the correction for a nonspherical shape. For aspect ratios less than 2, $a_2\sim1$ and can be neglected. We adopt $p=2$, $\Delta V=0.9$ km~s$^{-1}$ from NRO 45-m observations of the N$_2$H$^+$ J=1-0 line \citep{Kim2020}, and $R=7000$ au as derived from JCMT data. The resulting virial mass is 3.4 M$_\odot$, which exceeds the total gas mass of the core. However, a substantial fraction of the gas in the core has already been converted into stellar mass. The total stellar mass estimated by \cite{2023AJ....165..209R} is approximately 6.3 M$_\odot$. Thus, the combined gas and stellar mass of the core is about 8.6 M$_\odot$, which is significantly larger than the virial mass, indicating that the core remains gravitationally bound.

\section{Molecular outflows} \label{sec:outflows}
The ALMA CO (2-1) data used in this work have twice the angular resolution of those used in \cite{2023AJ....165..209R}, enabling a more detailed investigation of the molecular outflows. As shown in Figure \ref{fig:outflows}, all four protostars drive outflows. The gas mass of each outflow is estimated using the CO J=2-1 emission \citep{Qiu2009}:
\begin{equation}
M_{\rm out}=1.39\times10^{-6}\exp\left(\frac{16.59}{T_{\rm ex}}\right)(T_{\rm ex}+0.92)D^{2}\int\frac{\tau_{12}}{1-e^{-\tau_{12}}}S_{\nu}dv,
\end{equation}
where $M_{\rm out}$, $T_{\rm ex}$, $D$, $\tau_{12}$, and $S_{\nu}$ are the outflow gas mass in $M_{\odot}$, the excitation temperature of $^{12}$CO (2-1), the source distance in kpc, the optical depth of $^{12}$CO (2-1), and the line flux in Jy, respectively. The mass of each outflow lobe is presented in the fifth column of Table \ref{tab:outflows}. 

The mass-loss rate ($\dot{M}_{\rm loss}$) of the outflow is calculated via $\dot{M}_{\rm loss}=M_{\rm out}/t_{\rm dyn}$. The dynamical timescale of the outflow is defined as $t_{\rm dyn}=L/V_{\rm char}$, where $L$ denotes the length of the outflow lobe, and $V_{\rm char}$ represents the characteristic outflow velocity derived from the intensity-weighted velocity map. Assuming that the jet and wind energy originate from the gravitational energy released by mass accretion onto the protostar \citep{Bontemps1996}, the outflow force ($F_{\rm out}$) is related to the mass accretion rate ($\dot{M}_{\rm acc}$) through the following equation derived from momentum conservation:
\begin{equation}\label{eq_finalmdot}
\dot{M}_{\rm acc} = \frac{1}{f_{\rm ent}} \, \frac{\dot{M}_{\rm acc}}{\dot{M}_w} \frac {1}{V_w} \, F_{\rm flow},
\end{equation}
where we adopt a typical jet/wind velocity $V_w \sim 150$ km~s$^{-1}$ \citep{Bontemps1996}, $\dot{M}_w$ is the wind/jet mass-loss rate. Models of jet/wind formation predict, on average, $\dot{M}_w / \dot{M}_{\rm acc} \sim 0.1$ \citep{Shu1994,Pelletier1992,Wardle1993,Bontemps1996}. The entrainment efficiency is typically $f_{\rm ent} \sim 0.1-0.25$; we adopt $f_{\rm ent}=0.25$ here. The outflow force $F_{\rm flow}$ is calculated as:
\begin{equation}
F_{\rm flow}=\frac{P_{\rm flow}}{t_{\rm dyn}}=\frac{M_{\rm out}V_{\rm char}}{t_{\rm dyn}}=\dot{M}_{\rm out}V_{\rm char}
\end{equation}
The derived mass accretion rates are listed in the eighth column of Table \ref{tab:outflows}.

\begin{table*}[!thb]
\centering
\caption{Observed and derived outflow parameters for each outflow lobe.}
\renewcommand{\arraystretch}{1.5} 
\begin{tabular}{cccccccccc}
\hline
\hline
Outflow lobes         & $V_{\rm char}$ & $S_{\nu}$ & $L$ & $M_{\rm out}$ & $t_{\rm dyn}$ & $\dot{M}_{\rm loss}$ & $\dot{M}_{\rm acc}$\\
                      &  (km~s$^{-1}$) & (Jy~km~s$^{-1}$) & (au) & ($10^{-6}$ M$_{\odot}$) & (yr) &($10^{-8}$ M$_{\odot}$~yr$^{-1}$) &($10^{-8}$ M$_{\odot}$~yr$^{-1}$) \\
\hline
N2-M1 red    & 7.1  & 1.0 & 1200 & 12.2 & 800 & 1.5 & 2.9 \\
N2-M1 blue   & 4.4  & 0.7 & 800  & 8.5  & 860 & 1.0 & 1.2 \\
\hline
N2-M2 red    & 8.4  & 3.0 & 900  & 36.6 & 510 & 7.2 & 16.1 \\
N2-M2 blue   & 6.1  & 2.2 & 800  & 26.8 & 620 & 4.3 & 7.0 \\
\hline
N2-W1 red    & 9.4  & 0.3 & 250  & 3.7  & 130 & 2.9 & 7.3 \\
N2-W1 blue   & 4.8  & 0.8 & 400  & 9.8  & 400 & 2.5 & 3.2 \\
\hline
N2-W2 red    & 10.4 & 4.0 & 1400 & 48.8 & 640 & 7.6 & 21.2 \\
N2-W2 blue   & 9.9  & 5.1 & 1400 & 62.2 & 670 & 9.3 & 24.5 \\
\hline
\end{tabular}
\label{tab:outflows}
\end{table*}

\section{Differential rotation} \label{sec:rotation}
We use moment 1 maps of H$_2$CO lines to determine the velocity gradients in the central flattened core that hosts the two binary systems. To enhance sensitivity, we stacked data from three H$_2$CO J=3-2 line transitions. Figure \ref{fig:moment1} presents the moment maps and position-velocity cuts along the dashed lines. 

The velocity gradients ($\zeta=|\nabla v_{\rm lsr}|$) were determined by fitting the position-velocity cuts. The $\zeta$ values derived from ACA data, ACA+TM2 data, and ACA+TM1+TM2 data are approximately 37, 71, and 120 km~s$^{-1}$~pc$^{-1}$, respectively, indicating that the gas rotates faster in the inner region. The angular velocity $\omega$ can be calculated as $\omega = \zeta / \sin i$, where $i$ is the inclination of $\omega$ relative to the line of sight. Following \cite{Goodman1993}, the ratio of rotational kinetic energy to gravitational energy is given by:
\begin{equation}
\beta = \frac{(1/2)I\omega^2}{qGM^2/R} = \frac{1}{2}\frac{p}{q}\frac{\omega^2R^3}{GM}.
\end{equation}
We adopt a total gas mass of $M = 0.95$ M$_\odot$ and a radius of $R = 2400$ au for the central region containing N2-M and N2-W derived from ACA observations, and $p/q = 0.22$ for an $r^{-2}$ density profile. For $\zeta$ values of 37, 71, and 120 km~s$^{-1}$~pc$^{-1}$, the corresponding $\beta \sin^2 i$ values are 0.06, 0.21, and 0.61, respectively. For a uniform density profile, the $\beta$ value would increase by a factor of three. In any case, the core is likely undergoing fast differential rotation.

Interestingly, recent studies have also found evidence for velocity gradients across the filament that hosts the G205.46-14.56 core \citep{Hsieh2021}. The filament rotates more slowly, with a ratio of rotational energy to gravitational energy of $\beta \sim 0.04$, comparable to that of the rotating core observed with ACA.

\begin{figure*}[ht]
\centering
\includegraphics[scale=0.9]{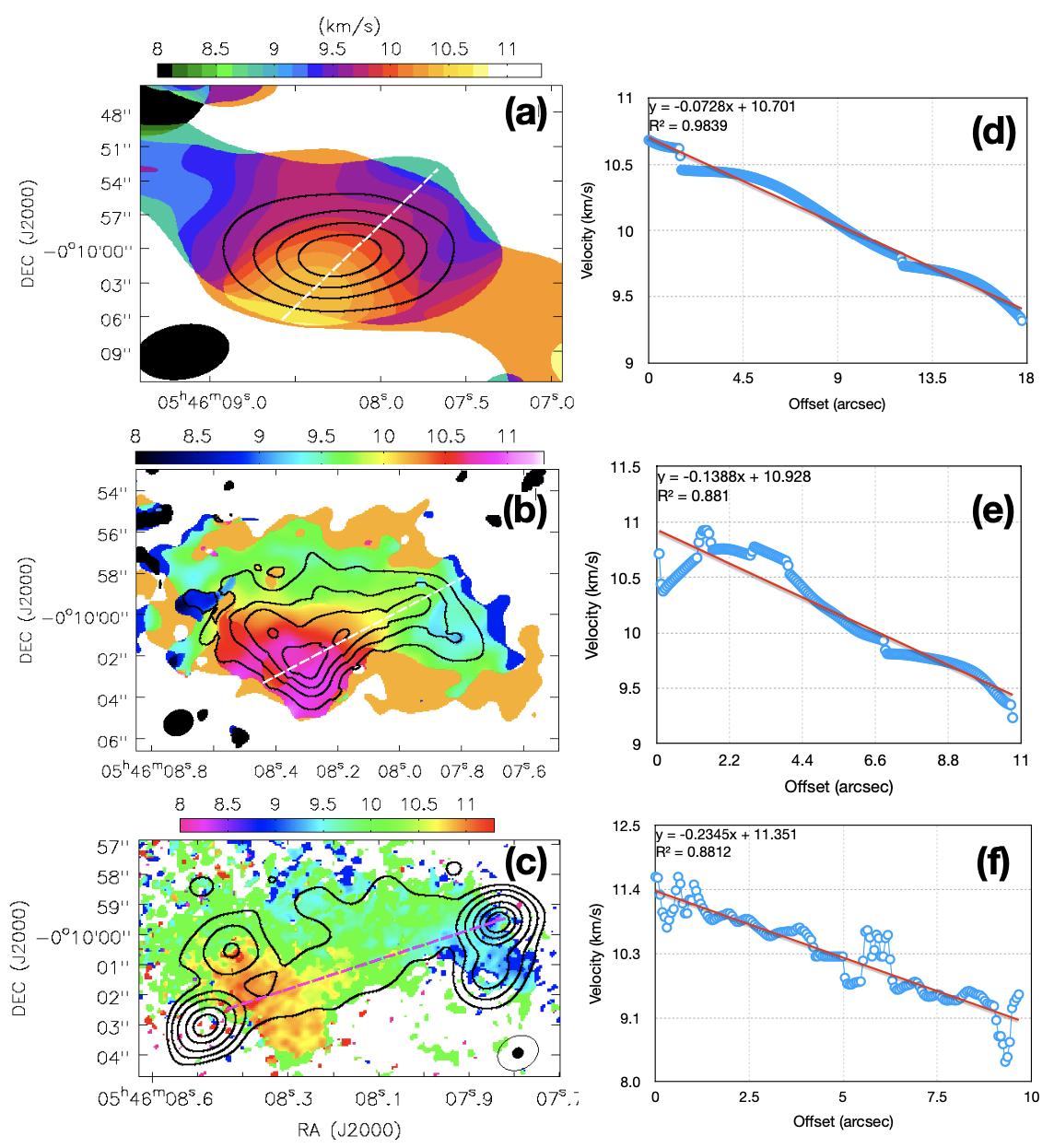}
\caption{(a) Moment 1 map of H$_2$CO line emission from ACA observations shown in color. The corresponding integrated intensity map is shown in contours. Contour levels are [0.2, 0.4, 0.6, 0.8]$\times$7.97 Jy~beam$^{-1}$~km~s$^{-1}$. (b) Moment 1 map of H$_2$CO line emission from ACA+TM2 observations shown in color. Contours show the integrated intensity map. Contour levels are [0.2, 0.4, 0.6, 0.8]$\times$0.398 Jy~beam$^{-1}$~km~s$^{-1}$. (c) Moment 1 map of H$_2$CO line emission from ACA+TM2+TM1 observations shown in color. Contours show the 1.3 mm continuum emission. Contour levels are [0.05, 0.1, 0.2, 0.4, 0.6, 0.8]$\times$42.7 mJy~beam$^{-1}$. Panels (d), (e), and (f) show the velocity profiles along the dashed lines in the corresponding left panels. Red lines indicate linear fits.}
\label{fig:moment1}
\end{figure*}

\section{Formation of the filament} \label{cloud}

Fig.~\ref{fig:PMO}a presents the large-scale moment 1 map of $^{13}$CO J=1-0 line emission. It shows a velocity gradient from southeast to northwest. The velocity gradient along the dashed line is $\sim0.4$ km~s$^{-1}$~pc$^{-1}$ (Fig.~\ref{fig:PMO}b). The dense filament G205.46-14.56 is located at the interface between the redshifted and blueshifted components of the cloud gas, as traced by the $^{13}$CO J=1-0 line emission (Fig.~\ref{fig:PMO}c). This spatial coincidence suggests that self-gravitating filaments may originate from large-scale converging flow within the molecular cloud \citep{Chen2020} or from cloud-cloud collision \citep{2021PASJ...73S.256E}.

\begin{figure*}[ht]
\centering
\includegraphics[scale=0.7]{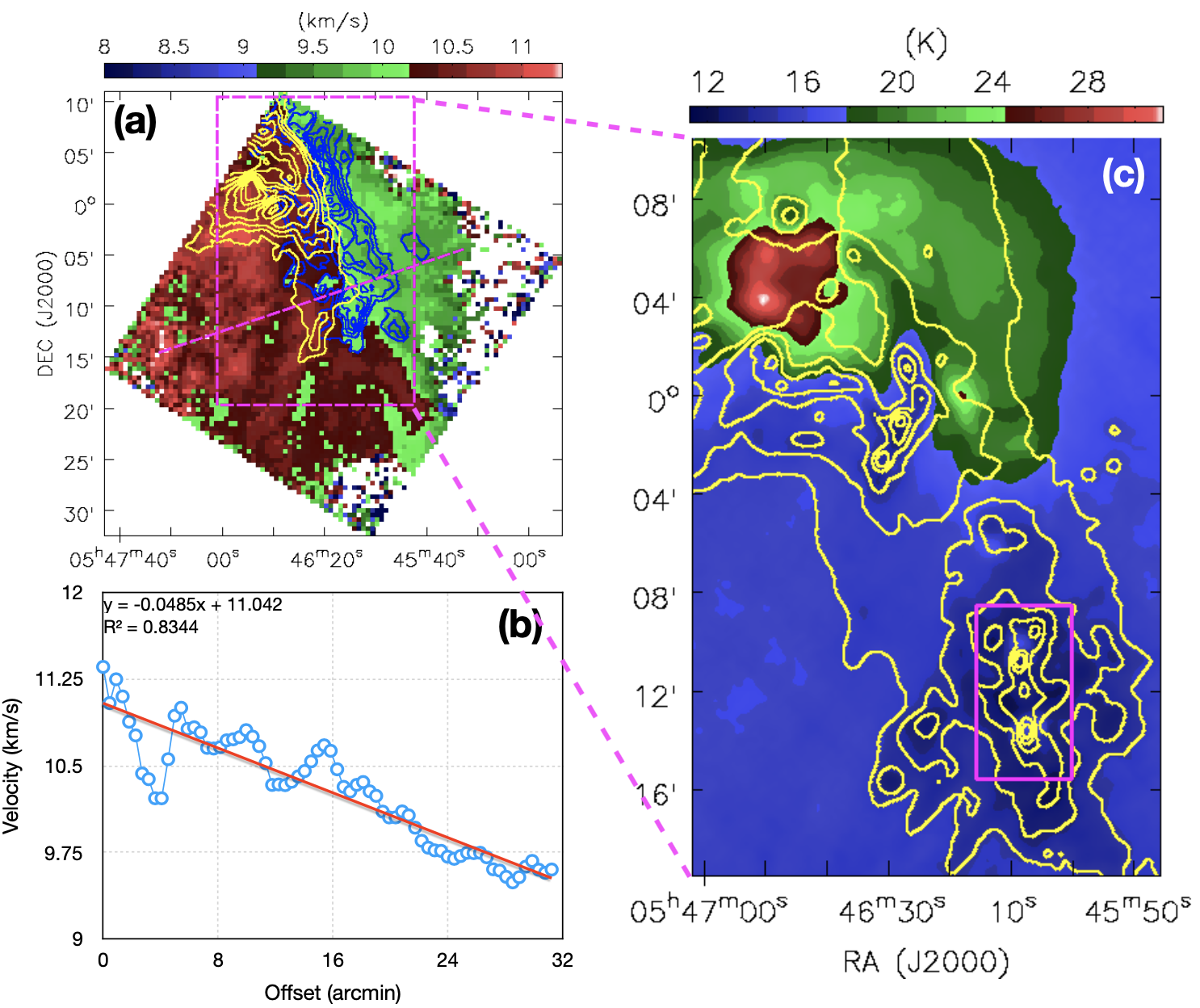}
\caption{(a) Moment 1 map of $^{13}$CO (1-0) in the velocity range 6--14 km~s$^{-1}$, shown in color scale. Redshifted emission integrated from 10.5 to 14 km~s$^{-1}$ is shown in yellow contours. Contour levels are [0.3, 0.4, 0.5, 0.6, 0.7, 0.8, 0.9]$\times$25.6 K~km~s$^{-1}$. Blueshifted emission integrated from 6 to 9.5 km~s$^{-1}$ is shown in blue contours. Contour levels are [0.3, 0.4, 0.5, 0.6, 0.7, 0.8, 0.9]$\times$10.1 K~km~s$^{-1}$. (b) Velocity profile along the dashed line shown in panel (a). The red line is a linear fit. (c) Dust temperature and column density maps derived from Herschel data by \cite{Konyves2020}. The color image shows the dust temperature map. Contours show the column density map. Contour levels are [0.01, 0.1, 0.2, 0.4, 0.6, 0.8]$\times$5.74$\times10^{22}$ cm$^{-2}$. The pink box marks the G205.46-14.56 filament.}
\label{fig:PMO}
\end{figure*}

\clearpage

\clearpage
\bibliography{sample}{}

@ARTICLE{2011ApJ...735...64W,
       author = {{Wang}, Ke and {Zhang}, Qizhou and {Wu}, Yuefang and {Zhang}, Huawei},
        title = "{Hierarchical Fragmentation and Jet-like Outflows in IRDC G28.34+0.06: A Growing Massive Protostar Cluster}",
      journal = {\apj},
         year = 2011,
        month = jul,
       volume = {735},
       number = {1},
          eid = {64},
        pages = {64},
          doi = {10.1088/0004-637X/735/1/64},
archivePrefix = {arXiv},
       eprint = {1105.4559},
 primaryClass = {astro-ph.GA},
       adsurl = {https://ui.adsabs.harvard.edu/abs/2011ApJ...735...64W}
}

@ARTICLE{2014MNRAS.439.3275W,
       author = {{Wang}, Ke and {Zhang}, Qizhou and {Testi}, Leonardo and {van der Tak}, Floris and {Wu}, Yuefang and {Zhang}, Huawei and {Pillai}, Thushara and {Wyrowski}, Friedrich and {Carey}, Sean and {Ragan}, Sarah E. and {Henning}, Thomas},
        title = "{Hierarchical fragmentation and differential star formation in the Galactic `Snake': infrared dark cloud G11.11-0.12}",
      journal = {\mnras},
         year = 2014,
        month = apr,
       volume = {439},
       number = {4},
        pages = {3275-3293},
          doi = {10.1093/mnras/stu127},
archivePrefix = {arXiv},
       eprint = {1401.4157},
 primaryClass = {astro-ph.GA},
       adsurl = {https://ui.adsabs.harvard.edu/abs/2014MNRAS.439.3275W}
}

@article{Gower1969,
 ISSN = {00359254, 14679876},
 URL = {http://www.jstor.org/stable/2346439},
 author = {J. C. Gower and G. J. S. Ross},
 journal = {Journal of the Royal Statistical Society. Series C (Applied Statistics)},
 number = {1},
 pages = {54--64},
 publisher = {[Wiley, Royal Statistical Society]},
 title = {Minimum Spanning Trees and Single Linkage Cluster Analysis},
 urldate = {2026-06-16},
 volume = {18},
 year = {1969}
}

@ARTICLE{2021PASJ...73S.256E,
       author = {{Enokiya}, Rei and {Ohama}, Akio and {Yamada}, Rin and {Sano}, Hidetoshi and {Fujita}, Shinji and {Hayashi}, Katsuhiro and {Tsutsumi}, Daichi and {Torii}, Kazufumi and {Nishimura}, Atsushi and {Konishi}, Ryotaro and {Yamamoto}, Hiroaki and {Tachihara}, Kengo and {Hasegawa}, Yutaka and {Kimura}, Kimihiro and {Ogawa}, Hideo and {Fukui}, Yasuo},
        title = "{High-mass star formation in Orion B triggered by cloud-cloud collision: Merging molecular clouds in NGC 2024}",
      journal = {\pasj},
         year = 2021,
        month = jan,
       volume = {73},
        pages = {S256-S272},
          doi = {10.1093/pasj/psaa049},
archivePrefix = {arXiv},
       eprint = {1912.11607},
 primaryClass = {astro-ph.GA},
       adsurl = {https://ui.adsabs.harvard.edu/abs/2021PASJ...73S.256E}
}

@ARTICLE{2014ApJ...794...44B,
       author = {{Boss}, Alan P. and {Keiser}, Sandra A.},
        title = "{Collapse and Fragmentation of Magnetic Molecular Cloud Cores with the Enzo AMR MHD Code. II. Prolate and Oblate Cores}",
      journal = {\apj},
         year = 2014,
        month = oct,
       volume = {794},
       number = {1},
          eid = {44},
        pages = {44},
          doi = {10.1088/0004-637X/794/1/44},
archivePrefix = {arXiv},
       eprint = {1408.2479},
 primaryClass = {astro-ph.SR},
       adsurl = {https://ui.adsabs.harvard.edu/abs/2014ApJ...794...44B}
}

@ARTICLE{2016ApJ...826..153L,
       author = {{Lim}, Jeremy and {Yeung}, Paul K.~H. and {Hanawa}, Tomoyuki and {Takakuwa}, Shigehisa and {Matsumoto}, Tomoaki and {Saigo}, Kazuya},
        title = "{Rotationally driven Fragmentation in the Formation of the Binary Protostellar System L1551 IRS 5}",
      journal = {\apj},
         year = 2016,
        month = aug,
       volume = {826},
       number = {2},
          eid = {153},
        pages = {153},
          doi = {10.3847/0004-637X/826/2/153},
archivePrefix = {arXiv},
       eprint = {1607.00323},
 primaryClass = {astro-ph.SR},
       adsurl = {https://ui.adsabs.harvard.edu/abs/2016ApJ...826..153L}
}

@ARTICLE{2018MNRAS.478.5460R,
       author = {{Riaz}, R. and {Vanaverbeke}, S. and {Schleicher}, D.~R.~G.},
        title = "{Formation of multiple low-mass stars, brown dwarfs, and planemos via gravitational collapse}",
      journal = {\mnras},
         year = 2018,
        month = aug,
       volume = {478},
       number = {4},
        pages = {5460-5472},
          doi = {10.1093/mnras/sty1409},
archivePrefix = {arXiv},
       eprint = {1805.09881},
 primaryClass = {astro-ph.SR},
       adsurl = {https://ui.adsabs.harvard.edu/abs/2018MNRAS.478.5460R}
}

@INPROCEEDINGS{2023ASPC..534..275O,
       author = {{Offner}, S.~S.~R. and {Moe}, M. and {Kratter}, K.~M. and {Sadavoy}, S.~I. and {Jensen}, E.~L.~N. and {Tobin}, J.~J.},
        title = "{The Origin and Evolution of Multiple Star Systems}",
    booktitle = {Protostars and Planets VII},
         year = 2023,
       editor = {{Inutsuka}, S. and {Aikawa}, Y. and {Muto}, T. and {Tomida}, K. and {Tamura}, M.},
       series = {Astronomical Society of the Pacific Conference Series},
       volume = {534},
        month = jul,
        pages = {275},
          doi = {10.48550/arXiv.2203.10066},
archivePrefix = {arXiv},
       eprint = {2203.10066},
 primaryClass = {astro-ph.SR},
       adsurl = {https://ui.adsabs.harvard.edu/abs/2023ASPC..534..275O}
}

@ARTICLE{2023A&A...673A.134M,
       author = {{Mignon-Risse}, R. and {Gonz{\'a}lez}, M. and {Commer{\c{c}}on}, B.},
        title = "{The role of magnetic fields in the formation of multiple massive stars}",
      journal = {\aap},
         year = 2023,
        month = may,
       volume = {673},
          eid = {A134},
        pages = {A134},
          doi = {10.1051/0004-6361/202345845},
archivePrefix = {arXiv},
       eprint = {2303.04528},
 primaryClass = {astro-ph.SR},
       adsurl = {https://ui.adsabs.harvard.edu/abs/2023A&A...673A.134M}
}

@ARTICLE{1991Natur.351..298B,
       author = {{Boss}, Alan P.},
        title = "{Formation of hierarchical multiple protostellar cores}",
      journal = {\nat},
         year = 1991,
        month = may,
       volume = {351},
       number = {6324},
        pages = {298-300},
          doi = {10.1038/351298a0},
       adsurl = {https://ui.adsabs.harvard.edu/abs/1991Natur.351..298B}
}

@ARTICLE{2023AJ....165..209R,
       author = {{Reipurth}, Bo and {Bally}, J. and {Yen}, Hsi-Wei and {Arce}, H.~G. and {Rodr{\'\i}guez}, L.-F. and {Raga}, A.~C. and {Geballe}, T.~R. and {Rao}, R. and {Comer{\'o}n}, F. and {Mikkola}, S. and {Aspin}, C.~A. and {Walawender}, J.},
        title = "{The HH 24 Complex: Jets, Multiple Star Formation, and Orphaned Protostars}",
      journal = {\aj},
         year = 2023,
        month = may,
       volume = {165},
       number = {5},
          eid = {209},
        pages = {209},
          doi = {10.3847/1538-3881/acadd4},
archivePrefix = {arXiv},
       eprint = {2301.01813},
 primaryClass = {astro-ph.SR},
       adsurl = {https://ui.adsabs.harvard.edu/abs/2023AJ....165..209R}
}

@ARTICLE{2022ApJ...931..158L,
       author = {{Luo}, Qiu-yi and {Liu}, Tie and {Tatematsu}, Ken'ichi and {Liu}, Sheng-Yuan and {Li}, Pak Shing and {di Francesco}, James and {Johnstone}, Doug and {Goldsmith}, Paul F. and {Dutta}, Somnath and {Hirano}, Naomi and {Lee}, Chin-Fei and {Li}, Di and {Kim}, Kee-Tae and {Won Lee}, Chang and {Lee}, Jeong-Eun and {Liu}, Xun-chuan and {Juvela}, Mika and {He}, Jinhua and {Qin}, Sheng-Li and {Liu}, Hong-Li and {Eden}, David and {Kwon}, Woojin and {Sahu}, Dipen and {Li}, Shanghuo and {Xu}, Feng-Wei and {Zhang}, Si-ju and {Hsu}, Shih-Ying and {Bronfman}, Leonardo and {Sanhueza}, Patricio and {Pelkonen}, Veli-Matti and {Zhou}, Jian-wen and {Liu}, Rong and {Gu}, Qi-lao and {Wu}, Yue-fang and {Mai}, Xiao-feng and {Falgarone}, Edith and {Shen}, Zhi-Qiang},
        title = "{ALMA Survey of Orion Planck Galactic Cold Clumps (ALMASOP): How Do Dense Core Properties Affect the Multiplicity of Protostars?}",
      journal = {\apj},
         year = 2022,
        month = jun,
       volume = {931},
       number = {2},
          eid = {158},
        pages = {158},
          doi = {10.3847/1538-4357/ac66d9},
archivePrefix = {arXiv},
       eprint = {2204.06176},
 primaryClass = {astro-ph.GA},
       adsurl = {https://ui.adsabs.harvard.edu/abs/2022ApJ...931..158L}
}

@ARTICLE{Goodman1993,
       author = {{Goodman}, A.~A. and {Benson}, P.~J. and {Fuller}, G.~A. and {Myers}, P.~C.},
        title = "{Dense Cores in Dark Clouds. VIII. Velocity Gradients}",
      journal = {\apj},
         year = 1993,
        month = apr,
       volume = {406},
        pages = {528},
          doi = {10.1086/172465},
       adsurl = {https://ui.adsabs.harvard.edu/abs/1993ApJ...406..528G}
}

@ARTICLE{Andre2010,
       author = {{Andr{\'e}}, Ph. and {Men'shchikov}, A. and {Bontemps}, S. and {K{\"o}nyves}, V. and {Motte}, F. and {Schneider}, N. and {Didelon}, P. and {Minier}, V. and {Saraceno}, P. and {Ward-Thompson}, D. and {di Francesco}, J. and {White}, G. and {Molinari}, S. and {Testi}, L. and {Abergel}, A. and {Griffin}, M. and {Henning}, Th. and {Royer}, P. and {Mer{\'\i}n}, B. and {Vavrek}, R. and {Attard}, M. and {Arzoumanian}, D. and {Wilson}, C.~D. and {Ade}, P. and {Aussel}, H. and {Baluteau}, J. -P. and {Benedettini}, M. and {Bernard}, J. -Ph. and {Blommaert}, J.~A.~D.~L. and {Cambr{\'e}sy}, L. and {Cox}, P. and {di Giorgio}, A. and {Hargrave}, P. and {Hennemann}, M. and {Huang}, M. and {Kirk}, J. and {Krause}, O. and {Launhardt}, R. and {Leeks}, S. and {Le Pennec}, J. and {Li}, J.~Z. and {Martin}, P.~G. and {Maury}, A. and {Olofsson}, G. and {Omont}, A. and {Peretto}, N. and {Pezzuto}, S. and {Prusti}, T. and {Roussel}, H. and {Russeil}, D. and {Sauvage}, M. and {Sibthorpe}, B. and {Sicilia-Aguilar}, A. and {Spinoglio}, L. and {Waelkens}, C. and {Woodcraft}, A. and {Zavagno}, A.},
        title = "{From filamentary clouds to prestellar cores to the stellar IMF: Initial highlights from the Herschel Gould Belt Survey}",
      journal = {\aap},
         year = 2010,
        month = jul,
       volume = {518},
          eid = {L102},
        pages = {L102},
          doi = {10.1051/0004-6361/201014666},
archivePrefix = {arXiv},
       eprint = {1005.2618},
 primaryClass = {astro-ph.GA},
       adsurl = {https://ui.adsabs.harvard.edu/abs/2010A&A...518L.102A}
}

@ARTICLE{Pineda2015,
       author = {{Pineda}, Jaime E. and {Offner}, Stella S.~R. and {Parker}, Richard J. and {Arce}, H{\'e}ctor G. and {Goodman}, Alyssa A. and {Caselli}, Paola and {Fuller}, Gary A. and {Bourke}, Tyler L. and {Corder}, Stuartt A.},
        title = "{The formation of a quadruple star system with wide separation}",
      journal = {\nat},
         year = 2015,
        month = feb,
       volume = {518},
       number = {7538},
        pages = {213-215},
          doi = {10.1038/nature14166},
       adsurl = {https://ui.adsabs.harvard.edu/abs/2015Natur.518..213P}
}

@ARTICLE{LeeJE2017,
       author = {{Lee}, Jeong-Eun and {Lee}, Seokho and {Dunham}, Michael M. and {Tatematsu}, Ken'ichi and {Choi}, Minho and {Bergin}, Edwin A. and {Evans}, Neal J.},
        title = "{Formation of wide binaries by turbulent fragmentation}",
      journal = {Nature Astronomy},
         year = 2017,
        month = aug,
       volume = {1},
          eid = {0172},
        pages = {0172},
          doi = {10.1038/s41550-017-0172},
archivePrefix = {arXiv},
       eprint = {1707.00233},
 primaryClass = {astro-ph.SR},
       adsurl = {https://ui.adsabs.harvard.edu/abs/2017NatAs...1E.172L}
}

@ARTICLE{Tobin2016b,
       author = {{Tobin}, John J. and {Kratter}, Kaitlin M. and {Persson}, Magnus V. and {Looney}, Leslie W. and {Dunham}, Michael M. and {Segura-Cox}, Dominique and {Li}, Zhi-Yun and {Chandler}, Claire J. and {Sadavoy}, Sarah I. and {Harris}, Robert J. and {Melis}, Carl and {P{\'e}rez}, Laura M.},
        title = "{A triple protostar system formed via fragmentation of a gravitationally unstable disk}",
      journal = {\nat},
         year = 2016,
        month = oct,
       volume = {538},
       number = {7626},
        pages = {483-486},
          doi = {10.1038/nature20094},
archivePrefix = {arXiv},
       eprint = {1610.08524},
 primaryClass = {astro-ph.SR},
       adsurl = {https://ui.adsabs.harvard.edu/abs/2016Natur.538..483T}
}

@ARTICLE{Tobin2016a,
       author = {{Tobin}, John J. and {Looney}, Leslie W. and {Li}, Zhi-Yun and {Chandler}, Claire J. and {Dunham}, Michael M. and {Segura-Cox}, Dominique and {Sadavoy}, Sarah I. and {Melis}, Carl and {Harris}, Robert J. and {Kratter}, Kaitlin and {Perez}, Laura},
        title = "{The VLA Nascent Disk and Multiplicity Survey of Perseus Protostars (VANDAM). II. Multiplicity of Protostars in the Perseus Molecular Cloud}",
      journal = {\apj},
         year = 2016,
        month = feb,
       volume = {818},
       number = {1},
          eid = {73},
        pages = {73},
          doi = {10.3847/0004-637X/818/1/73},
archivePrefix = {arXiv},
       eprint = {1601.00692},
 primaryClass = {astro-ph.SR},
       adsurl = {https://ui.adsabs.harvard.edu/abs/2016ApJ...818...73T}
}

@ARTICLE{Offner2010,
       author = {{Offner}, Stella S.~R. and {Kratter}, Kaitlin M. and {Matzner}, Christopher D. and {Krumholz}, Mark R. and {Klein}, Richard I.},
        title = "{The Formation of Low-mass Binary Star Systems Via Turbulent Fragmentation}",
      journal = {\apj},
         year = 2010,
        month = dec,
       volume = {725},
       number = {2},
        pages = {1485-1494},
          doi = {10.1088/0004-637X/725/2/1485},
archivePrefix = {arXiv},
       eprint = {1010.3702},
 primaryClass = {astro-ph.SR},
       adsurl = {https://ui.adsabs.harvard.edu/abs/2010ApJ...725.1485O}
}

@ARTICLE{Duchene2013,
       author = {{Duch{\^e}ne}, Gaspard and {Kraus}, Adam},
        title = "{Stellar Multiplicity}",
      journal = {\araa},
         year = 2013,
        month = aug,
       volume = {51},
       number = {1},
        pages = {269-310},
          doi = {10.1146/annurev-astro-081710-102602},
archivePrefix = {arXiv},
       eprint = {1303.3028},
 primaryClass = {astro-ph.SR},
       adsurl = {https://ui.adsabs.harvard.edu/abs/2013ARA&A..51..269D}
}

@ARTICLE{Hsieh2021,
       author = {{Hsieh}, Cheng-Han and {Arce}, H{\'e}ctor G. and {Mardones}, Diego and {Kong}, Shuo and {Plunkett}, Adele},
        title = "{Rotating Filament in Orion B: Do Cores Inherit Their Angular Momentum from Their Parent Filament?}",
      journal = {\apj},
         year = 2021,
        month = feb,
       volume = {908},
       number = {1},
          eid = {92},
        pages = {92},
          doi = {10.3847/1538-4357/abd034},
archivePrefix = {arXiv},
       eprint = {2012.02442},
 primaryClass = {astro-ph.GA},
       adsurl = {https://ui.adsabs.harvard.edu/abs/2021ApJ...908...92H}
}

@ARTICLE{Chen2020,
       author = {{Chen}, Che-Yu and {Mundy}, Lee G. and {Ostriker}, Eve C. and {Storm}, Shaye and {Dhabal}, Arnab},
        title = "{Self-gravitating filament formation from shocked flows: velocity gradients across filaments}",
      journal = {\mnras},
         year = 2020,
        month = may,
       volume = {494},
       number = {3},
        pages = {3675-3685},
          doi = {10.1093/mnras/staa960},
archivePrefix = {arXiv},
       eprint = {2004.02898},
 primaryClass = {astro-ph.GA},
       adsurl = {https://ui.adsabs.harvard.edu/abs/2020MNRAS.494.3675C}
}

@ARTICLE{Wardle1993,
       author = {{Wardle}, Mark and {Koenigl}, Arieh},
        title = "{The Structure of Protostellar Accretion Disks and the Origin of Bipolar Flows}",
      journal = {\apj},
         year = 1993,
        month = jun,
       volume = {410},
        pages = {218},
          doi = {10.1086/172739},
       adsurl = {https://ui.adsabs.harvard.edu/abs/1993ApJ...410..218W}
}

@ARTICLE{Pelletier1992,
       author = {{Pelletier}, Guy and {Pudritz}, Ralph E.},
        title = "{Hydromagnetic Disk Winds in Young Stellar Objects and Active Galactic Nuclei}",
      journal = {\apj},
         year = 1992,
        month = jul,
       volume = {394},
        pages = {117},
          doi = {10.1086/171565},
       adsurl = {https://ui.adsabs.harvard.edu/abs/1992ApJ...394..117P}
}

@ARTICLE{Shu1994,
       author = {{Shu}, Frank and {Najita}, Joan and {Ostriker}, Eve and {Wilkin}, Frank and {Ruden}, Steven and {Lizano}, Susana},
        title = "{Magnetocentrifugally Driven Flows from Young Stars and Disks. I. A Generalized Model}",
      journal = {\apj},
         year = 1994,
        month = jul,
       volume = {429},
        pages = {781},
          doi = {10.1086/174363},
       adsurl = {https://ui.adsabs.harvard.edu/abs/1994ApJ...429..781S}
}

@ARTICLE{Bontemps1996,
       author = {{Bontemps}, S. and {Andre}, P. and {Terebey}, S. and {Cabrit}, S.},
        title = "{Evolution of outflow activity around low-mass embedded young stellar objects}",
      journal = {\aap},
         year = 1996,
        month = jul,
       volume = {311},
        pages = {858-872},
       adsurl = {https://ui.adsabs.harvard.edu/abs/1996A&A...311..858B}
}

@ARTICLE{Qiu2009,
       author = {{Qiu}, Keping and {Zhang}, Qizhou and {Wu}, Jingwen and {Chen}, Huei-Ru},
        title = "{Submillimeter Array Observations of the Molecular Outflow in High-Mass Star-Forming Region G240.31+0.07}",
      journal = {\apj},
         year = 2009,
        month = may,
       volume = {696},
       number = {1},
        pages = {66-74},
          doi = {10.1088/0004-637X/696/1/66},
archivePrefix = {arXiv},
       eprint = {0901.4179},
 primaryClass = {astro-ph.GA},
       adsurl = {https://ui.adsabs.harvard.edu/abs/2009ApJ...696...66Q}
}

@ARTICLE{Kim2020,
       author = {{Kim}, Gwanjeong and {Tatematsu}, Ken'ichi and {Liu}, Tie and {Yi}, Hee-Weon and {He}, Jinhua and {Hirano}, Naomi and {Liu}, Sheng-Yuan and {Choi}, Minho and {Sanhueza}, Patricio and {T{\'o}th}, L. Viktor and {Evans}, Neal J., II and {Feng}, Siyi and {Juvela}, Mika and {Kim}, Kee-Tae and {Vastel}, Charlotte and {Lee}, Jeong-Eun and {Nguyễn Lu'o'ng}, Quang and {Kang}, Miju and {Ristorcelli}, Isabelle and {Feh{\'e}r}, Orsolya and {Wu}, Yuefang and {Ohashi}, Satoshi and {Wang}, Ke and {Kandori}, Ryo and {Hirota}, Tomoya and {Sakai}, Takeshi and {Lu}, Xing and {Thompson}, Mark A. and {Fuller}, Gary A. and {Li}, Di and {Shinnaga}, Hiroko and {Kim}, Jungha},
        title = "{Molecular Cloud Cores with a High Deuterium Fraction: Nobeyama Single-pointing Survey}",
      journal = {\apjs},
         year = 2020,
        month = aug,
       volume = {249},
       number = {2},
          eid = {33},
        pages = {33},
          doi = {10.3847/1538-4365/aba746},
archivePrefix = {arXiv},
       eprint = {2007.12319},
 primaryClass = {astro-ph.GA},
       adsurl = {https://ui.adsabs.harvard.edu/abs/2020ApJS..249...33K}
}

@ARTICLE{Mairs2017,
       author = {{Mairs}, Steve and {Lane}, James and {Johnstone}, Doug and {Kirk}, Helen and {Lacaille}, Kevin and {Bower}, Geoffrey C. and {Bell}, Graham S. and {Graves}, Sarah and {Chapman}, Scott and {JCMT Transient Team}},
        title = "{The JCMT Transient Survey: Data Reduction and Calibration Methods}",
      journal = {\apj},
         year = 2017,
        month = jul,
       volume = {843},
       number = {1},
          eid = {55},
        pages = {55},
          doi = {10.3847/1538-4357/aa7844},
archivePrefix = {arXiv},
       eprint = {1706.01897},
 primaryClass = {astro-ph.IM},
       adsurl = {https://ui.adsabs.harvard.edu/abs/2017ApJ...843...55M}
}

@ARTICLE{Herczeg2017,
       author = {{Herczeg}, Gregory J. and {Johnstone}, Doug and {Mairs}, Steve and {Hatchell}, Jennifer and {Lee}, Jeong-Eun and {Bower}, Geoffrey C. and {Chen}, Huei-Ru Vivien and {Aikawa}, Yuri and {Yoo}, Hyunju and {Kang}, Sung-Ju and {Kang}, Miju and {Chen}, Wen-Ping and {Williams}, Jonathan P. and {Bae}, Jaehan and {Dunham}, Michael M. and {Vorobyov}, Eduard I. and {Zhu}, Zhaohuan and {Rao}, Ramprasad and {Kirk}, Helen and {Takahashi}, Satoko and {Morata}, Oscar and {Lacaille}, Kevin and {Lane}, James and {Pon}, Andy and {Scholz}, Aleks and {Samal}, Manash R. and {Bell}, Graham S. and {Graves}, Sarah and {Lee}, E. 'lisa M. and {Parsons}, Harriet and {He}, Yuxin and {Zhou}, Jianjun and {Kim}, Mi-Ryang and {Chapman}, Scott and {Drabek-Maunder}, Emily and {Chung}, Eun Jung and {Eyres}, Stewart P.~S. and {Forbrich}, Jan and {Hillenbrand}, Lynne A. and {Inutsuka}, Shu-ichiro and {Kim}, Gwanjeong and {Kim}, Kyoung Hee and {Kuan}, Yi-Jehng and {Kwon}, Woojin and {Lai}, Shih-Ping and {Lalchand}, Bhavana and {Lee}, Chang Won and {Lee}, Chin-Fei and {Long}, Feng and {Lyo}, A. -Ran and {Qian}, Lei and {Scicluna}, Peter and {Soam}, Archana and {Stamatellos}, Dimitris and {Takakuwa}, Shigehisa and {Tang}, Ya-Wen and {Wang}, Hongchi and {Wang}, Yiren},
        title = "{How Do Stars Gain Their Mass? A JCMT/SCUBA-2 Transient Survey of Protostars in Nearby Star-forming Regions}",
      journal = {\apj},
         year = 2017,
        month = nov,
       volume = {849},
       number = {1},
          eid = {43},
        pages = {43},
          doi = {10.3847/1538-4357/aa8b62},
archivePrefix = {arXiv},
       eprint = {1709.02052},
 primaryClass = {astro-ph.SR},
       adsurl = {https://ui.adsabs.harvard.edu/abs/2017ApJ...849...43H}
}

@ARTICLE{Konyves2020,
       author = {{K{\"o}nyves}, V. and {Andr{\'e}}, Ph. and {Arzoumanian}, D. and {Schneider}, N. and {Men'shchikov}, A. and {Bontemps}, S. and {Ladjelate}, B. and {Didelon}, P. and {Pezzuto}, S. and {Benedettini}, M. and {Bracco}, A. and {Di Francesco}, J. and {Goodwin}, S. and {Rygl}, K.~L.~J. and {Shimajiri}, Y. and {Spinoglio}, L. and {Ward-Thompson}, D. and {White}, G.~J.},
        title = "{Properties of the dense core population in Orion B as seen by the Herschel Gould Belt survey}",
      journal = {\aap},
         year = 2020,
        month = mar,
       volume = {635},
          eid = {A34},
        pages = {A34},
          doi = {10.1051/0004-6361/201834753},
archivePrefix = {arXiv},
       eprint = {1910.04053},
 primaryClass = {astro-ph.SR},
       adsurl = {https://ui.adsabs.harvard.edu/abs/2020A&A...635A..34K}
}

@ARTICLE{Dutta2020,
       author = {{Dutta}, Somnath and {Lee}, Chin-Fei and {Liu}, Tie and {Hirano}, Naomi and {Liu}, Sheng-Yuan and {Tatematsu}, Ken'ichi and {Kim}, Kee-Tae and {Shang}, Hsien and {Sahu}, Dipen and {Kim}, Gwanjeong and {Moraghan}, Anthony and {Jhan}, Kai-Syun and {Hsu}, Shih-Ying and {Evans}, Neal J. and {Johnstone}, Doug and {Ward-Thompson}, Derek and {Kuan}, Yi-Jehng and {Lee}, Chang Won and {Lee}, Jeong-Eun and {Traficante}, Alessio and {Juvela}, Mika and {Vastel}, Charlotte and {Zhang}, Qizhou and {Sanhueza}, Patricio and {Soam}, Archana and {Kwon}, Woojin and {Bronfman}, Leonardo and {Eden}, David and {Goldsmith}, Paul F. and {He}, Jinhua and {Wu}, Yuefang and {Pelkonen}, Veli-Matti and {Qin}, Sheng-Li and {Li}, Shanghuo and {Li}, Di},
        title = "{ALMA Survey of Orion Planck Galactic Cold Clumps (ALMASOP). II. Survey Overview: A First Look at 1.3 mm Continuum Maps and Molecular Outflows}",
      journal = {\apjs},
         year = 2020,
        month = dec,
       volume = {251},
       number = {2},
          eid = {20},
        pages = {20},
          doi = {10.3847/1538-4365/abba26},
archivePrefix = {arXiv},
       eprint = {2010.14507},
 primaryClass = {astro-ph.GA},
       adsurl = {https://ui.adsabs.harvard.edu/abs/2020ApJS..251...20D}
}

@ARTICLE{Rosolowsky2008,
       author = {{Rosolowsky}, E.~W. and {Pineda}, J.~E. and {Kauffmann}, J. and {Goodman}, A.~A.},
        title = "{Structural Analysis of Molecular Clouds: Dendrograms}",
      journal = {\apj},
         year = 2008,
        month = jun,
       volume = {679},
       number = {2},
        pages = {1338-1351},
          doi = {10.1086/587685},
archivePrefix = {arXiv},
       eprint = {0802.2944},
 primaryClass = {astro-ph},
       adsurl = {https://ui.adsabs.harvard.edu/abs/2008ApJ...679.1338R}
}

@ARTICLE{Kong2021,
       author = {{Kong}, Shuo and {Arce}, H{\'e}ctor G. and {Shirley}, Yancy and {Glasgow}, Colton},
        title = "{Evidence of Core Growth in the Dragon Infrared Dark Cloud: A Path for Massive Star Formation}",
      journal = {arXiv e-prints},
         year = 2021,
        month = mar,
          eid = {arXiv:2103.08697},
        pages = {arXiv:2103.08697},
archivePrefix = {arXiv},
       eprint = {2103.08697},
 primaryClass = {astro-ph.GA},
       adsurl = {https://ui.adsabs.harvard.edu/abs/2021arXiv210308697K}
}

@ARTICLE{Krieger2020,
       author = {{Krieger}, Nico and {Bolatto}, Alberto D. and {Koch}, Eric W. and {Leroy}, Adam K. and {Rosolowsky}, Erik and {Walter}, Fabian and {Wei{\ss}}, Axel and {Eden}, David J. and {Levy}, Rebecca C. and {Meier}, David S. and {Mills}, Elisabeth A.~C. and {Moore}, Toby and {Ott}, J{\"u}rgen and {Su}, Yang and {Veilleux}, Sylvain},
        title = "{The Turbulent Gas Structure in the Centers of NGC 253 and the Milky Way}",
      journal = {\apj},
         year = 2020,
        month = aug,
       volume = {899},
       number = {2},
          eid = {158},
        pages = {158},
          doi = {10.3847/1538-4357/aba903},
archivePrefix = {arXiv},
       eprint = {2008.02518},
 primaryClass = {astro-ph.GA},
       adsurl = {https://ui.adsabs.harvard.edu/abs/2020ApJ...899..158K}
}

@ARTICLE{Phiri2021,
       author = {{Phiri}, S.~P. and {Kirk}, J.~M. and {Ward-Thompson}, D. and {Sansom}, A.~E. and {Bendo}, G.~J.},
        title = "{ALMA $^{13}$CO(J = 1 - 0) observations of NGC 604 in M33: Physical properties of molecular clouds}",
      journal = {\mnras},
         year = 2021,
        month = may,
          doi = {10.1093/mnras/stab1251},
archivePrefix = {arXiv},
       eprint = {2104.14701},
 primaryClass = {astro-ph.GA},
       adsurl = {https://ui.adsabs.harvard.edu/abs/2021MNRAS.tmp.1232P}
}

@ARTICLE{Takemura2021,
       author = {{Takemura}, Hideaki and {Nakamura}, Fumitaka and {Kong}, Shuo and {Arce}, H{\'e}ctor G. and {Carpenter}, John M. and {Ossenkopf-Okada}, Volker and {Klessen}, Ralf and {Sanhueza}, Patricio and {Shimajiri}, Yoshito and {Tsukagoshi}, Takashi and {Kawabe}, Ryohei and {Ishii}, Shun and {Dobashi}, Kazuhito and {Shimoikura}, Tomomi and {Goldsmith}, Paul F. and {S{\'a}nchez-Monge}, {\'A}lvaro and {Kauffmann}, Jens and {Pillai}, Thushara G.~S. and {Padoan}, Paolo and {Ginsberg}, Adam and {Smith}, Rowan J. and {Bally}, John and {Mairs}, Steve and {Pineda}, Jaime E. and {Lis}, Dariusz C. and {Burkhart}, Blakesley and {Schilke}, Peter and {Chen}, Hope How-Huan and {Isella}, Andrea and {Friesen}, Rachel K. and {Goodman}, Alyssa A. and {Harper}, Doyal A.},
        title = "{The Core Mass Function in the Orion Nebula Cluster Region: What Determines the Final Stellar Masses?}",
      journal = {\apjl},
         year = 2021,
        month = mar,
       volume = {910},
       number = {1},
          eid = {L6},
        pages = {L6},
          doi = {10.3847/2041-8213/abe7dd},
archivePrefix = {arXiv},
       eprint = {2103.08527},
 primaryClass = {astro-ph.GA},
       adsurl = {https://ui.adsabs.harvard.edu/abs/2021ApJ...910L...6T}
}

@ARTICLE{Hatchfield2020,
       author = {{Hatchfield}, H. Perry and {Battersby}, Cara and {Keto}, Eric and {Walker}, Daniel and {Barnes}, Ashley and {Callanan}, Daniel and {Ginsburg}, Adam and {Henshaw}, Jonathan D. and {Kauffmann}, Jens and {Kruijssen}, J.~M. Diederik and {Longmore}, Steve N. and {Lu}, Xing and {Mills}, Elisabeth A.~C. and {Pillai}, Thushara and {Zhang}, Qizhou and {Bally}, John and {Butterfield}, Natalie and {Contreras}, Yanett A. and {Ho}, Luis C. and {Ott}, J{\"u}rgen and {Patel}, Nimesh and {Tolls}, Volker},
        title = "{CMZoom. II. Catalog of Compact Submillimeter Dust Continuum Sources in the Milky Way's Central Molecular Zone}",
      journal = {\apjs},
         year = 2020,
        month = nov,
       volume = {251},
       number = {1},
          eid = {14},
        pages = {14},
          doi = {10.3847/1538-4365/abb610},
archivePrefix = {arXiv},
       eprint = {2009.05052},
 primaryClass = {astro-ph.GA},
       adsurl = {https://ui.adsabs.harvard.edu/abs/2020ApJS..251...14H}
}

@ARTICLE{Lewis2021,
       author = {{Lewis}, John Arban and {Lada}, Charles J. and {Bieging}, John and {Kazarians}, Anoush and {Alves}, Jo{\~a}o and {Lombardi}, Marco},
        title = "{Probing the Cold Deep Depths of the California Molecular Cloud: The Icy Relationship between CO and Dust}",
      journal = {\apj},
         year = 2021,
        month = feb,
       volume = {908},
       number = {1},
          eid = {76},
        pages = {76},
          doi = {10.3847/1538-4357/abc41f},
archivePrefix = {arXiv},
       eprint = {2010.11968},
 primaryClass = {astro-ph.GA},
       adsurl = {https://ui.adsabs.harvard.edu/abs/2021ApJ...908...76L}
}

@ARTICLE{Ossenkopf1994,
       author = {{Ossenkopf}, V. and {Henning}, Th.},
        title = "{Dust opacities for protostellar cores.}",
      journal = {\aap},
         year = 1994,
        month = nov,
       volume = {291},
        pages = {943-959},
       adsurl = {https://ui.adsabs.harvard.edu/abs/1994A&A...291..943O}
}

@ARTICLE{Hildebrand1983,
       author = {{Hildebrand}, R.~H.},
        title = "{The determination of cloud masses and dust characteristics from submillimetre thermal emission.}",
      journal = {\qjras},
         year = 1983,
        month = sep,
       volume = {24},
        pages = {267-282},
       adsurl = {https://ui.adsabs.harvard.edu/abs/1983QJRAS..24..267H}
}

@ARTICLE{Clarke2019,
       author = {{Clarke}, S.~D. and {Williams}, G.~M. and {Ib{\'a}{\~n}ez-Mej{\'\i}a}, J.~C. and {Walch}, S.},
        title = "{Determining the presence of characteristic fragmentation length-scales in filaments}",
      journal = {\mnras},
         year = 2019,
        month = apr,
       volume = {484},
       number = {3},
        pages = {4024-4045},
          doi = {10.1093/mnras/stz248},
archivePrefix = {arXiv},
       eprint = {1901.06205},
 primaryClass = {astro-ph.GA},
       adsurl = {https://ui.adsabs.harvard.edu/abs/2019MNRAS.484.4024C}
}
\bibliographystyle{aasjournal}

\end{document}